\documentclass[11pt]{article}

\usepackage[margin = 1in]{geometry}
\usepackage{graphicx}
\usepackage{hyperref}
\usepackage{setspace}
\usepackage{indentfirst}
\usepackage{comment}
\usepackage{color}
\usepackage{natbib}
\usepackage{amsmath}
\usepackage{amssymb}
\usepackage{enumitem}
\usepackage{amsmath,amssymb,amsthm}
\usepackage{dsfont}
\usepackage{float}
\usepackage{hyperref}
\usepackage{mdframed}
\usepackage{graphicx}
\usepackage{comment}
\usepackage{caption}
\usepackage{subcaption}
\usepackage{booktabs, threeparttable}

\newcommand{\p}{\mathbb{P}}
\newcommand{\bbeta}{\boldsymbol{\beta}}

\begin{document}

\title{Statistical Assessments of Representational Reforms: A Case Study from Los Angeles}
\author{%
Sarah Cannon\thanks{Sarah Cannon and Evan T. R. Rosenman contributed equally to this work and are co-first authors. Rosenman is the corresponding author (\href{mailto:erosenman@cmc.edu}{erosenman@cmc.edu}).}\\
Mathematical Sciences Department\\
Claremont McKenna College\\
\href{mailto:scannon@cmc.edu}{scannon@cmc.edu}
\and
Evan T. R. Rosenman$^*$\\
Mathematical Sciences Department\\
Claremont McKenna College\\
\href{mailto:erosenman@cmc.edu}{erosenman@cmc.edu}
}
\date{}

\maketitle

\begin{abstract}
Electoral reforms -- including changes to district boundaries, electoral rules, and the size of elected bodies -- are a recurrent feature of American state and municipal politics. The representational impact of such reforms is challenging to assess before they are implemented, and even well-intentioned civic reformers are often surprised by the unintended consequences of changes to the political system. 

We develop a statistical framework for the prospective evaluation of electoral reforms by integrating several modes of analysis: descriptive statistics of the voter file, probabilistic race and ethnicity imputation, ecological inference, and redistricting simulations via ensembles. Together, these tools provide a unified assessment of how proposed reforms affect whose voices are ultimately heard in government. We apply this framework to reforms currently being considered for the Los Angeles City Council, including increasing the number of single-member districts, adopting ranked-choice voting, and introducing multimember districts elected by proportional ranked choice voting. We find that expanding the Council alone is unlikely to substantially improve representation for underrepresented communities. In contrast, eliminating low-turnout primary elections and adopting multimember districts produce larger improvements in representational equity. 

Our analyses were shared with L.A.'s Charter Reform Commission via public testimony in late 2025. Although motivated by Los Angeles, the framework is broadly applicable to evaluating prospective electoral reforms across American jurisdictions. 
\end{abstract}



\section{Introduction}\label{sec:intro}

Los Angeles, California is the second largest city in the United States, and a global hub for the arts and the entertainment industry. The city grew and diversified rapidly throughout the twentieth century, reaching a current population of around 3.9 million residents. L.A. is also highly diverse: the 2020 U.S. Census estimated its population to be 47\% Hispanic, 29\% non-Hispanic White, 12\% Asian, and 8\% Black, with the remaining individuals identifying as multiracial or some other race \citep[][see Supplement~\ref{sec:race_defns} for the definitions of these categories]{uscensus2020}.

L.A. is governed by an elected mayor along with a 15-member City Council. The Council size has been fixed since 1925 \citep{vein2026lacitycouncil}. Council members are elected from single-member districts across the city, each representing over 250,000 constituents. This ratio makes Los Angeles an outlier, with more residents per councilor than any other city among the ten largest in the United States, including New York, Chicago, Houston, Phoenix, and Philadelphia.

While the size and selection method for the Council has been controversial for decades, the impetus for reform emerged out of a 2022 scandal, in which an audio tape of a conversation between three City Council members and a labor leader was leaked anonymously to Reddit. The tape made headlines for an extensive series of racist comments made by the Councilors, as well as their discussion of plans to gerrymander the City Council map. While all three implicated Councilors were eventually ousted from the Council, reformers also demanded more fundamental changes in the way the city self-governs. The City Council first formed an Ad Hoc Committee on Government Reform in early 2023, and later convened a 13-member Charter Reform Commission to make recommendations for changes to the city charter \citep{zahniser2024chartercommission}. 

As part of this push for governance restructuring, several overlapping reform proposals were championed by academics, nonprofits, and civic groups. These include

\begin{itemize}
\item Increasing the number of single-member districts.
\item Switching to ranked choice voting within single-member districts, obviating the need for primary elections.
\item Adding multimember or at-large districts, with representatives selected via ranked choice voting. 
\item Eliminating the current system under which candidates who receive a majority of votes in the primary are elected outright, without advancing to the general election. As of early 2027, 10 of the 15 councilors will have been elected in the primary -- rather than the general election -- in their most recent race. 
\end{itemize}

With support from the Haynes Foundation, our team analyzed these reforms using a novel combination of voter file analysis, race imputation, ecological inference, and redistricting simulations. By combining Los Angeles voter file records with decennial U.S. Census data, we estimate the racial composition of the electorate and use graph-based sampling algorithms to generate thousands of plausible future districting plans, allowing us to evaluate how alternative reforms may affect representation. Our analysis primarily considers issues related to racial and ethnic representation, asking whose preferences have the opportunity to be reflected on the Council. We also consider other segments of the population with potential common interests, such as proxy measures for renter vs. homeowner status and for progressive vs. moderate political ideology. 

We find substantial inequities in the current electoral system, which particularly disadvantage Hispanic and Asian Angelenos in their capacity to elect representatives of their choosing. Moreover, our redistricting simulations suggest that increasing the size of the Council alone is unlikely to rectify these challenges. Our analysis also suggests that representational improvements may be derived from eliminating primaries by using ranked choice voting in a single general election. Even greater improvements are possible through multimember districting plans, with Council members selected with proportional ranked choice voting. To the extent we advocate for expanding the Council, we recommend doing so via a multimember district model. Our results were shared with the Charter Reform Commission in October 2025. The Commission ultimately recommended expanding the Council to 25 members, elected from single-member districts, and to adopt ranked choice voting, but the Council delayed citywide consideration of these reforms until at least 2028 \citep{gomez2026panel}. 

While the reform process is ongoing in Los Angeles, the analysis methods used in this manuscript can be deployed in other contexts across the United States. Such analyses are especially valuable given the wave of mid-decade redistricting efforts underway across the United States, as well as the evolving legal landscape surrounding voting rights after the Supreme Court's \emph{Louisiana v. Callais} ruling, which substantially weakened the Voting Rights Act.

\section{Inequities in the Current System}\label{sec:voting}

\subsection{Voter File Analysis}\label{sec:vf_analysis}

We begin with an analysis of racial and ethnic voting patterns in Los Angeles. To do so, we first obtain the L.A. voter file. This dataset contains a row for every registered voter in the city, including information such as their party registration, age, gender, and voting participation over recent primaries and general elections. We collect several ``vintages" of the voter file, including versions as of the 2020, 2022, and 2024 general elections. All voter files are sourced from L2, Inc., a leading national nonpartisan data vendor. 

Racial self-identification is not collected on California's voter files, so it must be probabilistically imputed. We utilize a popular method known as Bayesian Improved Surname Geocoding (BISG) \citep{imai2016improving}. This method sets a prior distribution for the race of individuals living within a given Census geography (we use Census blocks), based on decennial Census counts. The posterior distribution is obtained by combining these priors with estimated race-surname distributions, also obtained from the Census. BISG is widely used in social science to study the disparate racial impacts of key policies, in fields such as policing \citep{edwa:lee:espo:19}, lending \citep{zhan:18}, and eviction \citep{hepb:loui:desm:20}. Several improvements to BISG have been suggested, including relaxing key independence assumptions \citep{McCartan09092025} and calibrating predictions to known totals \cite{10.1093/jrsssa/qnaf003}. We utilize an improved version introduced in \cite{imai2022addressing}, which allows for incorporation of first and middle names into the predictions. Further details can be found in Supplement~\ref{sec:raceImputation}. 

We also seek to identify the voting coalitions that support elected leaders in Los Angeles. As individual-level vote choices are obfuscated via the secret ballot, we make use of ecological inference to estimate individual-level voter preferences from aggregate data. We link each voter to aggregated electoral results for City Council and citywide elections based on their voting precinct and vote method (in person vs. vote by mail). For the 2022 and 2024 elections, this yields approximately 1,600 to 2,000 unique ballot groupings across the city. We then deploy the method of \citep{fishman2024estimating} to estimate the probability that each voter supported the winning candidate in each election. Further details can be found in Supplement~\ref{sec:ecoreg}.

\subsection{Current Voting Patterns}\label{sec:current-voting-patterns}

\begin{figure}[H]
\centering

\subfloat[]{%
\includegraphics[width=.8\textwidth]{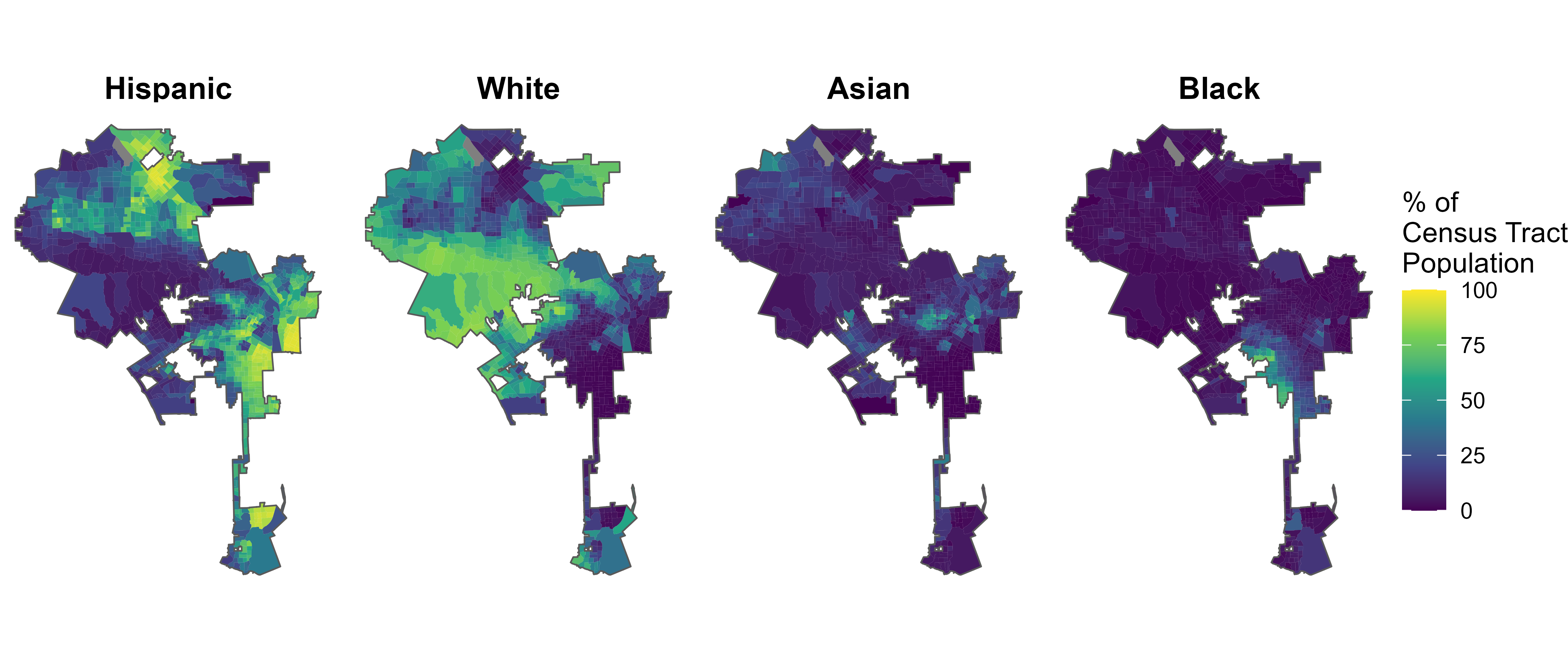}
\label{fig:MinorPopMap}
}

\vspace{0.5em}

\subfloat[]{%
\includegraphics[width=.8\textwidth]{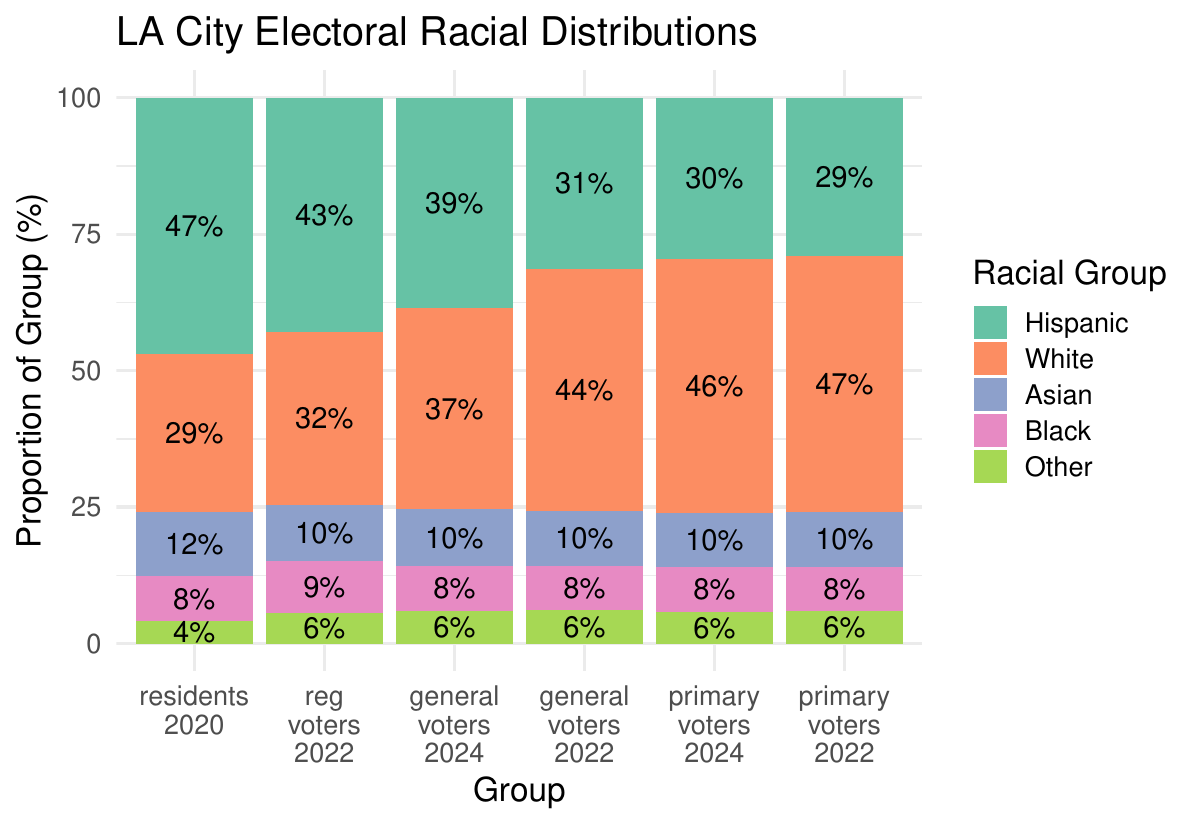}
\label{fig:LARacial}
}
\caption{(a) The distribution of Hispanic, White, Asian, and Black voters in Los Angeles. Each Census tract is colored according to the fraction of its residents who are in each racial group. (b) According to our BISG predictions the racial breakdown of L.A. residents, registered voters, and voters in various recent elections.}
\label{fig:LARaceSummary}
\end{figure}

We next turn to the question of who elects Los Angeles' Councilors and citywide officials. L.A. is a highly racially segregated city, as is visualized in Figure~\ref{fig:MinorPopMap}. White voters are concentrated on the city's West side, while Hispanic voters are clustered in East L.A. and in the city's northernmost stretches. Black voters are highly concentrated in South LA, while Asian residents (despite outnumbering Black residents) are spread more diffusely throughout the city.

This provides useful context to our next task: estimating the racial composition of electorates. Using the predicted probabilities from BISG, we compute the racial and ethnic distribution of registered voters as well as general election and primary voters in 2022, summarized in Figure \ref{fig:LARacial}. We find that the registered voter pool in Los Angeles is skewed toward White voters relative to the city's population. Moreover, in every election we studied, White voters comprise a larger portion of the electorate than their proportion of registered voters, largely at the expense of Hispanic voters. The White-Hispanic differential increases further in midterm election years and in primaries. Contrarily, Asian and Black voters comprise a relatively stable fraction of both the registered voter bloc and electorates in elections of all kinds. 

Both the diffuse geographic distribution of Asian Angelenos, and the differential participation trends between Hispanic and White Angelenos, directly manifest at the district level. We estimate that no district's registered voters pool constitutes more than 19\% Asian registered voters, and as a consequence, Asian voters never comprise more than 17.4\% of the electorate in any of the four elections we considered in 2022 and 2024. Moreover, in the current City Council districting plan, there are 7 districts that have a Hispanic majority when considering population and 6 majority-Hispanic districts by registered voters, but only 4 districts in which the 2022 general or primary electorate was majority-Hispanic. On the other hand, there are 3 districts with that are majority-White by population or by registered voters, but 5 that are majority-White among 2022 general election voters, and 6 among 2022 primary voters. For the full compositional breakdown by district, see Figure~\ref{fig:LARacialByDistrict} in Supplement~\ref{sec:distRaceComp}. 
Also see Figure~\ref{fig:15} which shows these results are not unique to the currently enacted districting plan.

\subsection{Who Selects L.A.'s Leaders?}\label{sec:who-selects-leaders}

\begin{table}[!t]
\caption{Demographic composition of Los Angeles and City Council electorates\label{tab:demo-electorates}}%
\begin{tabular*}{\columnwidth}{@{\extracolsep{\fill}}lccc@{\extracolsep{\fill}}}
\toprule
Demographic &
\shortstack{LA Population\\(2020)} &
\shortstack{City Council\\Electorates\\(2022, 2024)} &
\shortstack{\% of Voters\\Supporting Winning\\Councilor (2022, 2024)} \\
\midrule
Total    & 3.9M   & 869K & -- \\
Hispanic & 47\% & 33\%      & 56\% \\
White    & 29\% & 43\%      & 62\% \\
Asian    & 12\% & 10\%      & 52\% \\
Black    & 8\%  & 8\%       & 70\% \\
\bottomrule
\end{tabular*}
\begin{tablenotes}%
\item Population shares are computed using the total population from the 2020 Census. Electoral shares are estimated using BISG race probabilities computed from the 2022 and 2024 voter files for the primaries and general election voters for each year. 
\end{tablenotes}
\end{table}

Using these datasets, we are able to precisely characterize the set of voters who actually elect the City Council. Odd-numbered Council seat elections are held in midterm years, while even-numbered Council seat elections are held in presidential years. Hence, over the 2022 and 2024 cycles, every Council seat was up for election, and every seat was contested. Councilors elected from Districts 1, 3, 7, 9, 4, 6, 8, and 12 received the majority of the vote in their primaries, while the remaining seats were decided in the general election. 

In Table \ref{tab:demo-electorates}, we aggregate this information to identify the precise set of voters who selected the current City Council -- that is, the voters in each district who cast a ballot in the primary or general election in which the Councilor was elected. We call this group the ``Council-selecting electorate." We estimate that a small number of voters (approximately 10,000) appear more than once in these data, owing to relocations between elections. After deduplication, we find that approximately 870 thousand voters selected the current City Council. This population comprises about 40\% of registered voters in the city. Moreover, White voters are dramatically overrepresented in this group relative to their share of all Angelenos (43\% vs. 29\%) while Hispanic voters are dramatically underrepresented (33\% vs. 29\%). Asian voters are slightly underrepresented (12\% vs. 10\%), while Black voters comprise the same share of the Council-selecting electorate as of the population (8\%). 

The final column in Table  \ref{tab:demo-electorates} estimates the percentage of voters within each racial group in the Council-selecting electorate who actually supported the winning Councilor. These values are computed using our ecological inference estimates. While a majority of voters from each community supported winning Councilors, the values differ considerably across racial groups. Most notably, we see a large divergence on this measure between L.A.'s two smaller minority populations. 52\% of Asian voters in the Council-selecting electorate supported the winner, while the same was true of 70\% of Black voters in the Council-selecting electorate. 

This discrepancy is downstream of the geographic distribution of voters summarized in Figure \ref{fig:MinorPopMap}, along with the particulars of each community's preferences in the Council elections. We estimate that Asian voters comprise between 10\% and 17\% of the electorate in seven districts -- in decreasing order: Districts 1, 12, 10, 13, 14, 5, and 11 -- but that most Asian voters supported a losing candidate in District 1, 10, 13, and 11. Meanwhile, Black voters are heavily concentrated in Districts 8, 9, and 10, and we estimate that more than 80\% of Black voters in each district supported the winning Councillor. 

An important caveat to this analysis is that not all City Council seats are contested with equal vigor. If our goal is to establish that a community can translate its preferences into a voice in city governance, an uncompetitive election in 2022 or 2024 might mask a community's tendency to be outvoted by other, geographically proximal communities. Hence, as a complementary measure, we consider the results of a single, citywide race: the 2022 mayoral race. 

L.A.'s 2022 mayor's race is useful because it was a genuinely competitive election in which Congresswoman Karen Bass triumphed over real estate developer Rick Caruso by less than 10 percentage points. The race was contested in large part because both candidates were registered Democrats, with Caruso occupying the ``moderate" lane and Bass the ``progressive" lane. Bass won 10 City Council districts, while Caruso won five. We estimate each voter's support for Bass with a citywide ecological inference model, and report the percentage of each racial group's voters would have voted for the winning candidate within their City Council district in L.A.'s 2022 mayoral race. 

We find that 60\% of White voters, 56\% of Hispanic and Asian voters, and 73\% of Black voters supported the winner of the Bass vs. Caruso race within their own Council districts -- numbers that align quite closely with those in the final column of Table \ref{tab:demo-electorates}. This underscores the idea that L.A.'s racial communities are distributed with different levels of geographic ``efficiency" throughout the city.

\subsection{Discussion}\label{sec:geographic-inefficiency}

Our analyses in Section \ref{sec:who-selects-leaders} reflect a number of representational challenges in L.A.'s current system. L.A.'s system of automatically electing Councilors who receive a majority of the vote in the primary is especially problematic . Primary electorates are significantly less representative than those in general elections, meaning the subset of voters who choose L.A.'s leaders is particularly skewed. A straightforward alternative would be to advance the top two vote-getters in the primary to the general election, regardless of whether one exceeds 50\%. More fundamentally, the city could eliminate City Council primaries altogether and instead adopt ranked choice voting in the general election, allowing voters to rank candidates rather than select only one. 

The preceding patterns suggest that representational barriers differ across communities. For Hispanic voters, the central challenge is participation: lower registration and turnout, particularly in primaries, reduce the community's share of the decisive electorate. For Asian voters, the central challenge is geographic: even with substantial citywide population share, the community is more diffuse and is often outvoted by geographically proximate groups with different preferences.

\section{Why More Single-member Districts Do Not Solve the Problem}\label{sec:moredists}

\subsection{Council Expansion and District Sampling}\label{sec:council-expansion-sampling}

L.A. is an outlier in terms of the current ratio of residents to city councilors, with each councilor representing over 250,000 residents. This observation raises an important question: can expanding the number of City Council districts improve representation among communities of interest? One might posit that an increase in the number of seats could allow smaller enclaves of residents to be represented, making it easier for minority populations to attain representation.

We study the effects of increasing the number of single-member districts on the Council using an approach known as {\it ensemble analysis}. While we cannot know how districts will be drawn in the future, we can use computational tools to explore the space of possibilities.  First, we must merge data available at two different levels of geography (Census blocks and precincts) together into a unified data set; details of this process are provided in Supplement~\ref{sec:prorate}. 

To randomly sample a large number of potential districting plans, we used the Recombination Markov Chain \citep[ReCom;][]{DeFord2021Recombination}; technical details are provided in Supplement~\ref{sec:recom}. 
The method traverses the space of possible districting plans by repeatedly merging two districts and then randomly splitting this union into two new districts. This method enforces both compactness and population balance, two key metrics in any districting plan. Since our ensemble analyses cover so many potential districting plans, it allows us to observe what a ``natural" distribution of potential plans might look like, when only compactness and population balance are considered. 

\begin{figure}[ht]
\centering
\subcaptionbox{\label{fig:W15}}[.47\textwidth]{%
    \centering
    \includegraphics[width=\linewidth, trim=0 0 24bp 0, clip]{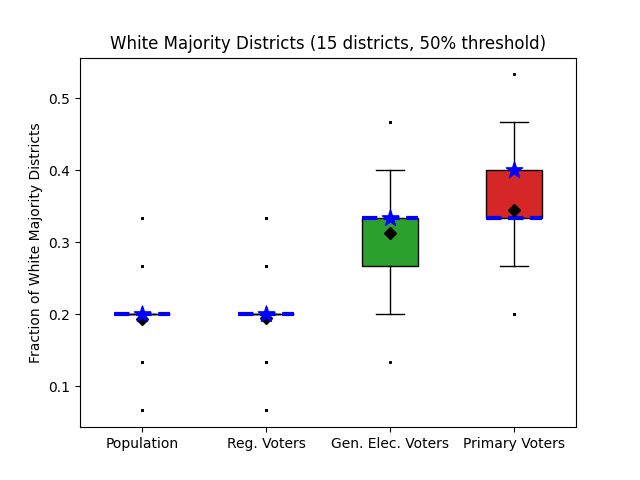}%
}\hfill
\subcaptionbox{}[.47\textwidth]{%
    \centering
    \includegraphics[width=\linewidth, trim=0 0 24bp 0, clip]{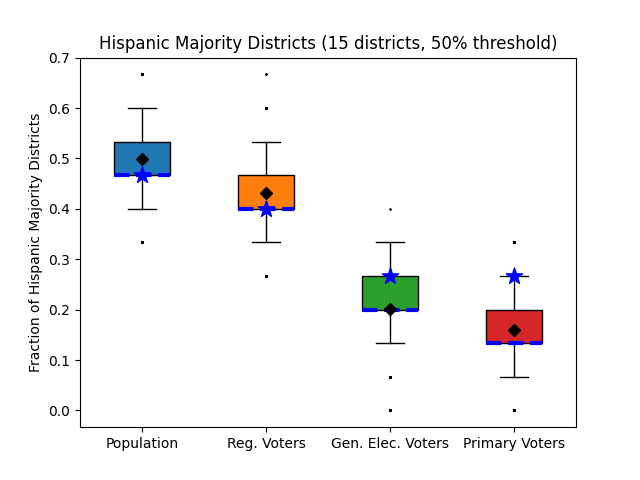}%
}

\caption{(a) Box plots showing, for randomly generated plans with 15 districts, the fraction of those 15 districts that are more than 50\% White according to population as well as 2022 registered voters, general election voters, and primary voters . The colored boxes are the middle 50\% of the random plans we generated in our ensemble, the blue lines are ensemble median values, the black diamonds are ensemble means, and the blue star is the currently enacted plan. (b) The same results for Hispanic-majority districts.
 }
\label{fig:15}
\end{figure}

To validate our methods, we first see where the currently enacted plan falls among our randomly sampled plans with 15 districts. Figure~\ref{fig:15} looks at the fraction of districts that have a White majority or a Hispanic majority across L.A.'s
population, 2022 registered voters, 2022 general election voters, and 2022 primary voters.  The colored boxes show the middle 50\% of our randomly sampled plans (i.e. the {\it interquartile range}), with sample means denoted as black diamonds, sample medians as blue dashed lines, and the values for the currently enacted plan as blue stars. The plot demonstrates the patterns observed in the enacted plan with regard to White and Hispanic representation are in fact typical of plans with 15 districts that only take compactness and population balance into account.

\subsection{Ensemble Results for Larger Councils}\label{sec:larger-councils}

\begin{figure}[H]
\centering

\subfloat[]{%
\begin{minipage}{.49\textwidth}
\centering
\includegraphics[width=\linewidth]{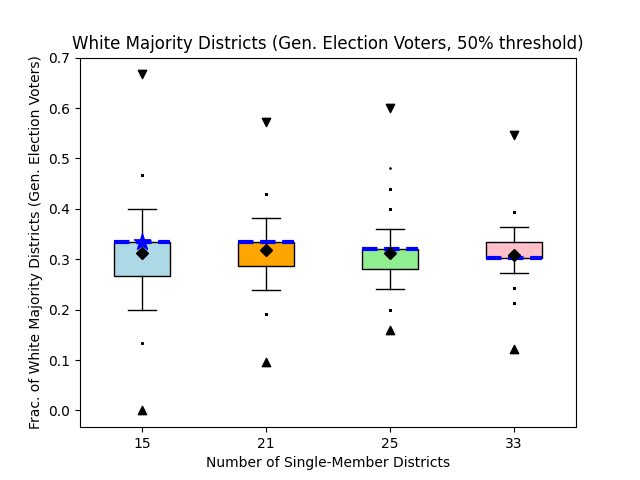}
\label{fig:WGEN}
\end{minipage}
}
\hfill
\subfloat[]{%
\begin{minipage}{.49\textwidth}
\centering
\includegraphics[width=\linewidth]{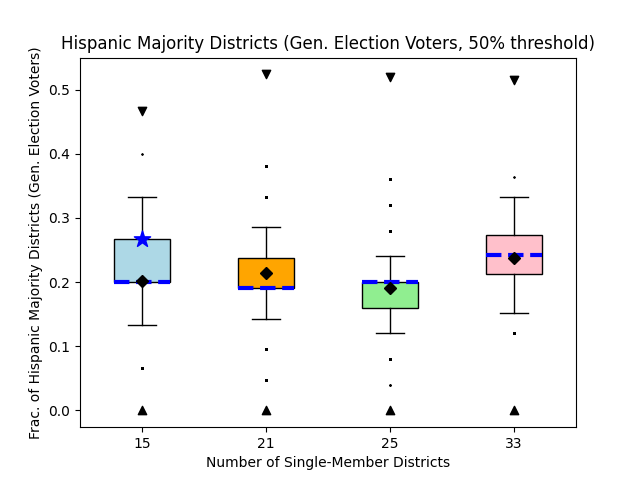}
\label{fig:HGEN}
\end{minipage}
}

\vspace{0.5em}

\subfloat[]{%
\begin{minipage}{.49\textwidth}
\centering
\includegraphics[width=\linewidth]{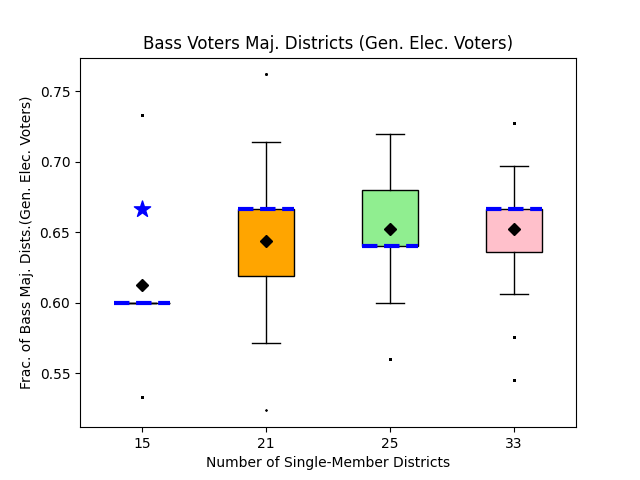}
\label{fig:Bass}
\end{minipage}
}
\hfill
\subfloat[]{%
\begin{minipage}{.49\textwidth}
\centering
\includegraphics[width=\linewidth]{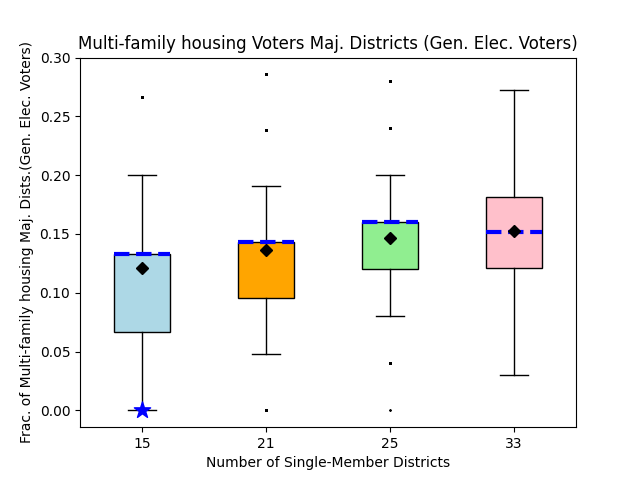}
\label{fig:MGEN}
\end{minipage}
}

\caption{
(a) Box plots showing the fraction of all districts where more than 50\% of general-election voters are White, for plans with 15, 21, 25, and 33 districts. The colored boxes represent the middle 50\% of plans in our ensemble, blue lines denote medians, black diamonds denote means, blue stars indicate the enacted plan, and triangles show extrema identified via short-burst optimization. Panels (b)--(d) present analogous results for Hispanic-majority districts, districts where Karen Bass received more votes than Rick Caruso, and districts where multifamily-housing residents outnumber single-family-housing residents.
}
\label{fig:W&HGEN}
\end{figure}

We run the same experiments for 15, 21, 25, and 33 districts, 
generating 300,000 plans for each. For convergence checks, we find that this is a sufficient number of plans; see Supplement~\ref{sec:mixing} 
for further details. 
In Figure~\ref{fig:W&HGEN}, we plot the fraction of districts that are White-majority and Hispanic-majority, when considering 2022 general election voters. 
The interquartile range, means, and medians are again denoted by colored boxes, blue dashed lines, and black diamonds, respectively. 
Importantly, there is little change in both plots across the different numbers of districts. In Figure~\ref{fig:WGEN}, for 15, 21, 25, and 33 districts, the interquartile range, mean, and median for White-majority districts remain essentially unchanged.  We see slightly more change in the interquartile ranges among  Hispanic-majority districts in Figure~\ref{fig:HGEN}, but there is still little movement in the means and medians. This demonstrates there is little change in the typical levels of racial representation when the number of districts grows. Additional plots for other groups and other thresholds (e.g. the fraction of districts that are more than 40\% Hispanic, Black-majority districts, etc.) are shown in Supplement~\ref{sec:more_single_plots}
in all, there was little change in typical representation when the number of districts increased.  

\subsection{Political Minorities and Multifamily Housing Residents}\label{sec:political-multifamily}

Our analysis did not solely consider representation through a racial lens.
In Figure~\ref{fig:Bass}, we consider the distribution of districts in which there were more votes for Karen Bass than Rick Caruso in L.A.'s 2022 Mayoral race. 
This analysis provides a rough proxy for the representation of progressive vs. moderate voters in Los Angeles. Meanwhile, in Figure~\ref{fig:MGEN}, we consider districts where more general election voters live in multifamily vs. single family housing, a rough proxy for renter status.\footnote{Voters were characterized as living in multifamily housing based on the presence of keywords -- Apt/Apartment, Unit, Suite/Ste, Building/Bldg, or Fl/Floor -- in their residential address, as reported on the voter file.} As our ensemble of sampled districting plans was made taking only compactness and equal population into account, and the enacted plan is an outlier with regard to Bass-won districts and majority-multifamily housing districts, 
it is likely that redistricting priorities other than compactness and equal population 
have led to an over-representation of Bass voters and under-representation of voters in multifamily housing in the currently enacted plan. Despite this, we still see relatively little change in the space of possibilities across the numbers of districts considered.

\subsection{Extremal Plans and the Range of Possible Outcomes}\label{sec:extremal-plans}

In addition to exploring the properties of typical districting plans drawn with only compactness and population balance in mind, we can also explore districting plans drawn to accomplish other goals.  Using a method known as {\it short bursts}~\citep{shortbursts}, we can create districting plans that aim to maximize or minimize a certain quality, such as the number of Hispanic-majority districts (see Supplement~\ref{sec:sb} 
for more details).  This gives us information on what could happen if a future redistricting commission chose to take such an approach, by exploring 
the frontier of what is possible. For example, in Figure~\ref{fig:WGEN}, when districting plans are drawn with consideration for only compactness and population balance, 
most 15-district plans yield between 3 and 6 majority-White districts (20-40\%); 
however, as the triangles in this figure show, it is possible to have as few as 0 and as many as 10 (67\%) of the districts as majority-White. Having tighter bounds is usually viewed as favorable because it provides less room for potentially malicious actors to derail the districting process. 
Considering the bounds provided by the triangles in Figures~\ref{fig:WGEN} and~\ref{fig:HGEN},
we see the more districts there are, the narrower the bounds for White voters and the larger the bounds for Hispanic voters, though these changes are slight. 
Overall, we find that increasing the number of districts without other reforms produces little change in the delegation of power.

\subsection{Single-member Ranked Choice Voting as a Primary-replacement Reform}
\label{sec:irv}

One widely championed reform -- ultimately proposed by the Charter Reform Commission itself -- 
is switching from plurality elections, in which the candidate with the most votes wins, to elections in which voters rank candidates, and these rankings are combined in order to produce a winner. Such an approach could eliminate or reduce the influence of primary elections, in which voters are less representative of the city as a whole. In this section, we study what could happen in a future ranked choice election, and how those results interact with the number of districts.

As in Section \ref{sec:who-selects-leaders}, we use the 2022 mayoral race as the basis for our analysis. While, in reality, different candidates would be running in each district, the mayoral race's outcome within each district provides a useful proxy for electing a progressive candidate versus a moderate candidate, respectively. We simulated potential ranked-choice ballots based on both primary and general election voting data, generated a large number of districting plans, and identified which candidate would win in an Instant Runoff Voting (IRV) election in each district. 

While only Bass and Caruso advanced to the November general election, 12 candidates appeared on the ballot in the June primary. We simulate ranked-choice voting for a slate of the six most popular candidates: Mike Feuer, Karen Bass, and Gina Viola (which we called the {\it progressive slate}) and Kevin de León,\footnote{de León was one of the Councilors caught on tape in the 2022 scandal, but we are using results from the primary election, which occurred before the tapes were leaked.} Andrew Kim, and Rick Caruso (which we called the {\it moderate slate}).
Within each voting precinct, we used a well-known sampling method to generate sample ranked ballots where up to all six of these candidates are ranked \citep{Donnay2025}. See Supplement~\ref{sec:ranked_ballots} 
for technical details. 

Parameters such as relative rankings within each moderate or progressive slate, and the likelihood of cohesion in each slate, were determined in order to match voting data within each precinct. For example, if in a particular precinct Bass received twice as many general election votes as Caruso, then twice as many ballots would favor the progressive slate compared to the moderate slate. If in that same precinct, Bass received three times as many primary votes compared to Viola, then three times as many ballots favoring the progressive slate would list Bass before Viola. This allows us to reasonably extrapolate ranked ballots from our current election data, in which voters are asked to provide a single choice rather than a ranking. We conducted this analysis for each of the 900+ precincts in Los Angeles in the 2022 primary election.   

\begin{figure}[H]
    \centering
    \includegraphics[width=1\linewidth]{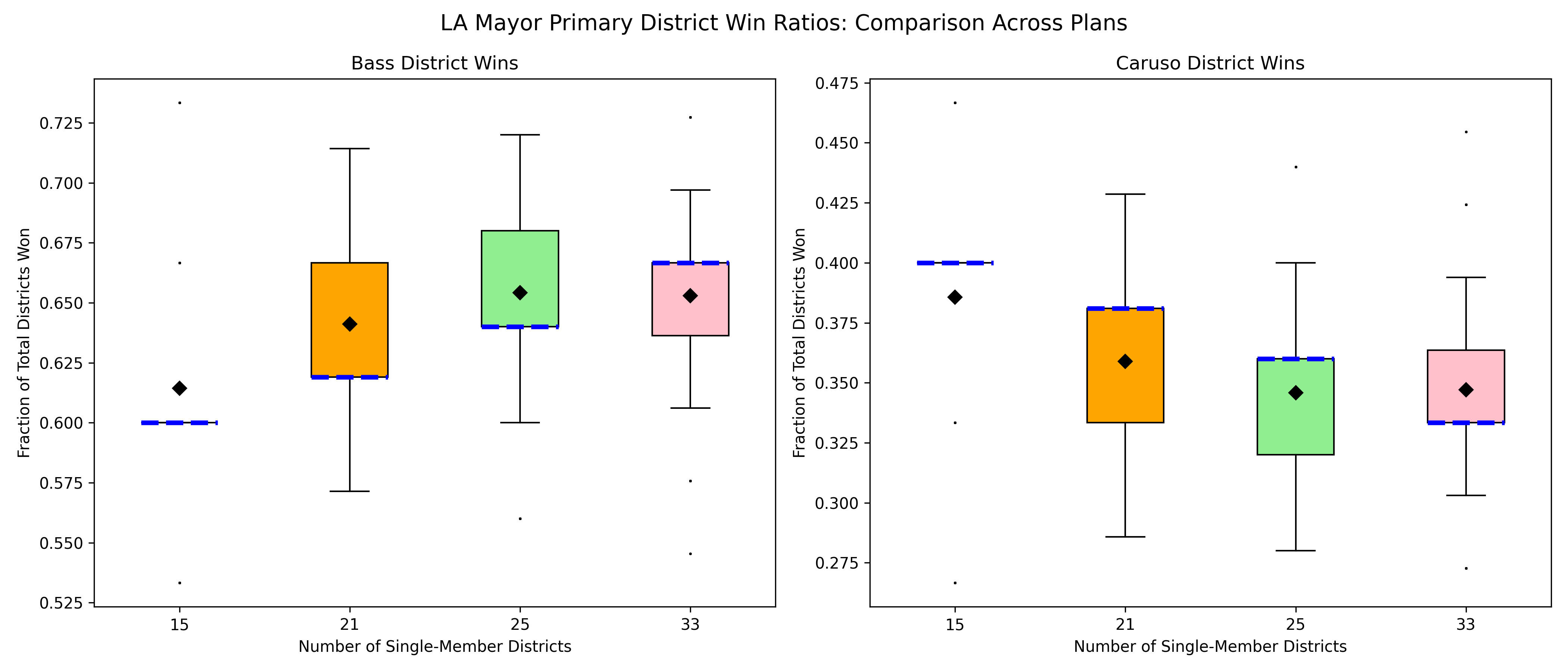}
    \caption{Distribution of districts won by Bass and Caruso in a 15-, 21-, 25-, and 33-district plan. The colored boxes represent the middle 50\% of the districts won per candidate, the black diamonds are the ensemble means, and the blue lines are the median values of districts won.}
    \label{fig:irv_results}
\end{figure}

We then combined these ballots with ensemble analysis; see Supplement~\ref{sec:irv-stv} 
for details. As above, we generated 300,000 redistricting plans of 15, 21, 25, and 33 districts.  Then, using our ranked ballots, we determined who would win in each district in an Instant Runoff Voting election. 
As shown in Figure~\ref{fig:irv_results}, we see a mild positive correlation between the number of single-member districts and the fraction of districts that would vote for Bass under our model. 
Furthermore, the fraction of districts that would elect Caruso in our model mildly decreases as the number of districts increases. 
All other candidates besides Caruso and Bass do not get enough votes to win even a single district, which is reflective of the results in the 2022 primary, in which Bass and Caruso combined to receive nearly 80\% of the vote, and no other candidate topped 8\%. 

Note the similarity to Figure~\ref{fig:Bass}, where we saw nearly the same pattern. 
This suggests that using instant runoff voting, at least in this case, wouldn't significantly change electoral outcomes, though it has the added benefit of 
rendering unnecessary a lower-turnout, less representative primary election.

\section{Proposal: Multimember Districts with Proportional Ranked Choice Voting}\label{sec:mmd}

In this section, we consider an alternate districting system: three-member districts with Council Members chosen via proportional ranked choice voting \citep[PRCV, also known as the Single Transferable Vote method; see e.g.][]{stephanopoulos2025ranked}. Under this system, voters in a given district would rank their preferred candidates, with votes reallocated from winners and eliminated candidates until three candidates surpass a threshold of $25\%$ of active ballots. A version of this system has long been used to elect the City Council in Cambridge, Massachusetts; PRCV was also recently adopted in Portland, Oregon and several smaller cities throughout the U.S. \citep{stephanopoulos2025ranked}. At the outset, the proposal carries with it several benefits. It obviates the need for primaries, which show poorer turnout for minority communities, especially Hispanic voters. It allows for non-majoritarian coalitions to elect Councilors of their choosing. It also reduces the number of Council districts (and conflicts in drawing their boundaries), while simultaneously allowing the Council to expand.

\subsection{Why Three-Member PRCV Districts Change the Threshold for Representation}\label{sec:mmd-ens}

We begin with a threshold-based analysis. In a three-member district with Council members elected using PRCV, if more than 25\% of a district's voters prefers the same candidate, that candidate will be elected (as opposed to the 50\% threshold for elections required in single-member districts). We use this threshold to assess what fraction of the council a minority group would have the opportunity to elect: for single-member districts, we count the number of districts in which a group forms a majority, and for three-member districts, we count the number of districts in which the group comprises more than 25\%, with districts where they surpass 50\% counted twice and districts where they surpass 75\% counted three times.

\begin{figure}[ht]
    \centering
    \includegraphics[width=0.65\linewidth]{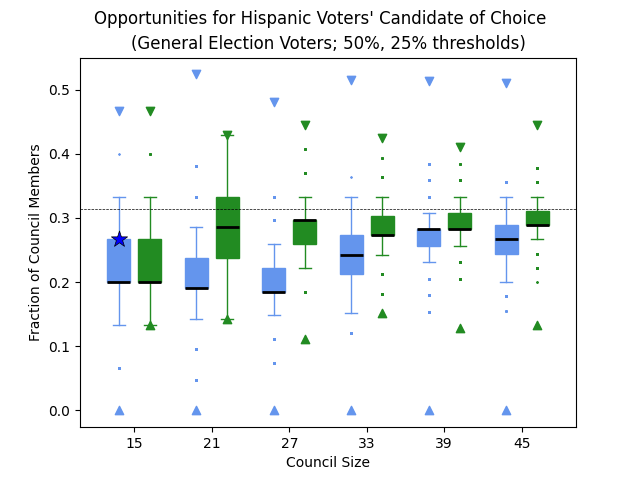}
    \caption{Proportion of Councilors that could be selected by Hispanic voters, considering Council sizes of 15, 21, 27, 33, 39, and 45, and plans with both single-member districts (in blue) and three-member districts (in green). Median values are given by thick horizontal black lines; maximum and minimum values obtained via the short bursts method are represented by triangles; and the horizontal dashed line stretching the length of the plot represents the overall fraction of 2022 general election voters that are Hispanic, a target if proportionality of election outcomes is a goal.}
    \label{fig:HGEN_50p_25p}
\end{figure}

As can be seen in Figure \ref{fig:HGEN_50p_25p}, increasing the number of members per district from one to three increases the number of candidates of the Hispanic community's choice in modal outcomes. In turn, this brings the percentage of candidates chosen by Hispanic voters closer to the percentage of 2022 Los Angeles general election voters that was Hispanic (represented by the dashed line). 

There is also typically a narrower range for the possible number of Hispanic-majority districts, as denoted by the triangles representing the maximum and minimum values observed, when employing the short bursts method to maximize or minimize Hispanic representation. The adoption of an independent commission to draw maps in the upcoming redistricting cycle provides some safeguard against extreme outcomes. As discussed above, this restricted range can be seen as an additional guardrail against bad actors, 
especially given the rapidly evolving legal landscape surrounding the Voting Rights Act and minority opportunity districts.

\begin{figure}[H]
    \centering
    \includegraphics[width=0.65\linewidth]{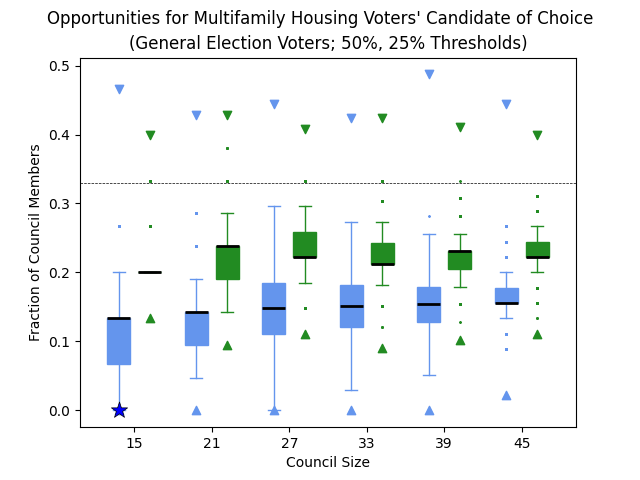}
    \caption{Proportion of councilors that could be selected by voters living in multifamily housing, considering Council sizes of 15, 21, 27, 33, 39, and 45, and plans with both single-member districts (in blue) and three-member districts (in green). Median values are given by thick horizontal black lines; maximum and minimum values obtained via the short bursts method are represented by triangles; and the horizontal dashed line stretching the length of the plot gives the overall fraction of 2022 general election voters that live in multifamily housing, a target if proportionality of election outcomes is a goal.}
    \label{fig:MGEN_50p_25p}
\end{figure}

We see a similar effect when considering the capacity of multifamily housing residents to elect Councilors of their choosing. In Figure~\ref{fig:MGEN_50p_25p}, we see that for any Council size, it is impossible to completely shut out multifamily housing voters from representation when using three-member districts, while such a lack of representation is possible under single-member districts for all council sizes up to 39 (as is the case in the currently enacted plan). Three-member districts also allow for closer-to-proportional representation for voters in multifamily housing when compared to single-member districts. 

Additional figures can be found in Supplement~\ref{sec:single_vs_multi_figs}
We find consistently that three-member districts induce a narrower range of possible representational outcomes for communities of interest, and that typical outcomes are closer to proportional representation for these communities. 

\subsection{Ranked Ballots and Single Transferrable Vote (STV) Elections}

In this section, rather than looking only at a vote threshold, we simulate ranked choice three-member elections and report the results. Ranked ballots are generated for each precinct following the same procedure as in Section \ref{sec:irv}, utilizing voting data from the 2022 primary and general mayoral elections. We focus on the 15-district case. 

\begin{figure}[ht]
    \centering
    \includegraphics[width=0.6\linewidth, trim = 25 0 0 0, clip]{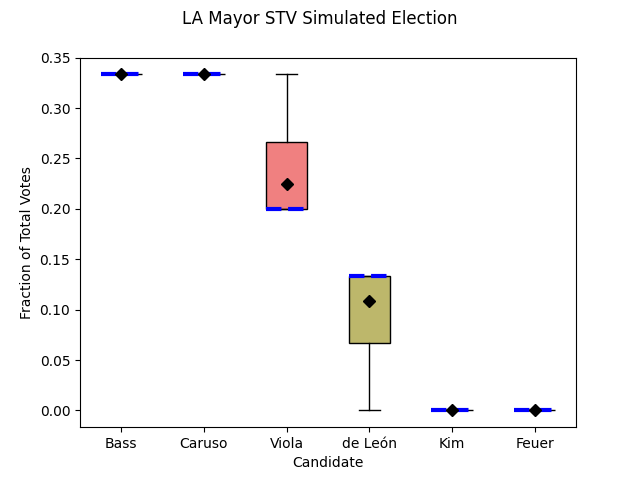}
    \caption{Distribution of candidates elected for simulated Single Transferrable Voting in five three-member districts, with ranked ballots generated based on votes in the 2022 primary and general Mayoral elections. The y-axis represents the fraction of the council comprised of theoretical future candidates with similar support as each named candidate. The dashed blue line gives the median of these results; the black diamond represents the mean; and the box represents the middle 50\% of the votes per candidate.}
    \label{fig:prcv_15}
\end{figure}

We simulated PRCV on a district plan of five members, with each district electing three candidates.
Again, we are using candidates in the 2022 mayoral primary
as proxies for future district-level elections, in which there will be different candidates in each district. 
To generate our sampled five-district plans, we ran 10 simultaneous runs of 30,000 steps each,  each from a different random starting state, aggregating the data for 300,000 total districting plans.\footnote{Computational and time constraints limited our ability to perform a single, ongoing run of 300,000 steps.}
Based on the results (see Figure~\ref{fig:prcv_15}), each district consistently elects both Bass and Caruso, giving voice to both the progressive and moderate voters in each district. The third candidate elected in each district varies, with Gina Viola typically chosen as the third Council member in three to four of the districts, and with Kevin de León elected as the third Council member in one to two  of the districts. This follows slate expectations from the 2022 primary election, in which more voters selected progressive candidates than moderates. 
These results demonstrate that no bloc completely dominates the election, yet the progressive bloc typically elects more candidates than the moderate bloc, as might be expected. Hence, switching to multimember districts with ranked choice voting is unlikely to result in significant changes to Council representation along the political spectrum, at least as represented by the candidates of the 2022 mayoral election. 

\section{Conclusion, Recommendations, and Next Steps}\label{sec:concl}

We address the potential representational effects of changes to the size of the City Council and the method in which Council Members are elected. While L.A.'s extreme Councilor-to-constituent ratio provides a compelling reason to expand the Council, we do not find that improved representation of minority communities would be a likely consequence of increasing the number of single-member districts. Rather, we find that minority communities have the opportunity to achieve roughly similar levels of representation across all possible council sizes.

However, we do recommend two reforms that have the potential to improve representation along multiple dimensions.  First -- as primary elections consistently see reduced participation from communities such as Hispanic voters and voters who live in multifamily housing -- we recommend that all Council member elections proceed to a general election regardless of whether a candidate exceeds 50\% of the primary vote. Alternately, switching to a system such as ranked choice voting obviates the need for primary elections, with voters instead ranking all candidates during a single general election. Finally, while requiring more dramatic structural change, a switch from single-member districts to three-member districts, with Council members elected via Proportional Ranked Choice Voting (PRCV), has the potential to improve electoral prospects for many minority communities by allowing for non-majoritarian coalitions to select Councilors of their choosing, allowing the election of candidates whose views more closely represent those of all Angelenos. 

As of summer 2026, the ultimate outcome of Los Angeles's charter reform effort remains uncertain. The results presented here were shared with the Charter Reform Commission in late 2025. In its report to the City Council in spring 2026, the Commission recommended expanding the Council to 25 seats and adopting ranked-choice voting, but did not recommend the use of multimember districts. The Council subsequently declined to place either reform on the 2026 ballot, opting instead for further study.

Regardless of the fate of these particular proposals, the analytic framework developed herein is straightforward and readily transferable to other cities and states. As debates over voting rights, representation, and partisan gerrymandering continue across the United States, quantitative methods can help clarify the tradeoffs inherent in alternative electoral systems.  The methods described herein provide a useful framework for civic reformers in that ongoing effort. 

\section*{Acknowledgments}
Many student research assistants participated in the project at various stages, ranging from initial data curation and cleaning to final polishing of figures included in this report. We gratefully acknowledge helpful contributions from: Ainslee Archibald, Quinten Carney, Danzhe Chen, Gabriel Dalton, Stella Cheng, Nicole David, Armine Kardashyan, Catherine Ma, Meghna Pamula, Derik Suria, Devon Xiong, and Katherine Zhao.

\section*{Funding}
This work was primarily supported by a generous planning grant from the Haynes Foundation, under their Governance and Democracy in the Los Angeles Region Initiative. Early preliminary work on this project was supported by NSF grant CCF-2104795. 
S. Cannon is also supported by NSF grant CCF-2443221 and by a Structural Democracy Fellowship from Cornell University, funded by the Crankstart Foundation. E. Rosenman is also supported by NSF grant DMS-2418829. 
Any opinions, findings, and conclusions or recommendations expressed in this material are those of the author(s) and do not necessarily reflect the views of the National Science Foundation.

\bibliographystyle{apalike}
\bibliography{la_bib}

\newpage

\section*{Supplementary Material}

\addcontentsline{toc}{section}{Supplement}
\renewcommand{\thesection}{\Alph{section}}
\renewcommand{\theHsection}{supplement.\Alph{section}}
\setcounter{section}{0}

\section{Definitions of Race and Ethnicity Used}\label{sec:race_defns}

Our initial race and ethnicity data comes from the 2020 U.S. Census, which separately asks respondents to self-identify with regard to race and ethnicity.  Respondents are asked to consider the following racial categories, and are able to select as many as apply: 
\begin{enumerate}
    \item White – A person having origins in any of the original peoples of Europe, the Middle East, or North Africa.
    \item Black or African American – A person having origins in any of the Black racial groups of Africa.
\item American Indian or Alaska Native – A person having origins in any of the original peoples of North and South America (including Central America) and who maintains tribal affiliation or community attachment.
\item Asian – A person having origins in any of the original peoples of the Far East, Southeast Asia, or the Indian subcontinent including, for example, Cambodia, China, India, Japan, Korea, Malaysia, Pakistan, the Philippine Islands, Thailand, and Vietnam.
\item Native Hawaiian or Other Pacific Islander – A person having origins in any of the original peoples of Hawaii, Guam, Samoa, or other Pacific Islands.
\item Some Other Race
\end{enumerate}
For ethnicity, respondents are separately asked if they are Hispanic or Latino or Not Hispanic or Latino. 

Given this census data, we grouped data into the five categories we consider (White, Black, Hispanic, Asian, and Other) following a standard procedure: 
\begin{enumerate}
    \item {\it White}: Total, Not Hispanic or Latino, Population of one race, White alone (census code P2\_005N)
    \item {\it Black}: Total, Not Hispanic or Latino, Population of one race, Black or African American alone (census code P2\_006N)
    \item {\it Hispanic}: Total, Hispanic or Latino (census code P2\_002N) 
    \item {\it Asian}: Total, Not Hispanic or Latino, Population of one race, Asian alone (census code P2\_008N) and Total, Not Hispanic or Latino, Population of one race, Native Hawaiian and Other Pacific Islander alone (census code P2\_009N). 
    \item {\it Other}: All other respondents; Total, Not Hispanic or Latino, Population of one race, American Indian and Alaska Native alone (census code P2\_007N) and Total, Not Hispanic or Latino, Population of one race, Some Other Race alone (census code P2\_010N) and Total, Not Hispanic or Latino, Population of two or more races (census code P2\_011N).    
\end{enumerate}
Our further race imputations are based on these five categories. 

\section{Race Imputation}\label{sec:raceImputation}

A key element of our analysis is race imputation, the probabilistic estimation of race and ethnicity in the absence of labeled data. The vast majority of U.S. states, including California, do not collect race and ethnicity data when citizens register to vote. Nevertheless, we have access to information about each voter's name, as well as the Census geography in which the voter lives, and hence can make informed predictions of each voter's self-identification. 

To generate these predictions, we deploy Bayesian Improved Surname Geocoding (BISG). This method sets a prior distribution for the race of individuals living within a given Census geography (typically, block or tract) based on the decennial Census counts of members of each racial group within the geography. The posterior distribution is then obtained by combining these priors with estimated race-surname distributions, also obtained from the Census. 

Formally, suppose there are $K$ possible racial categories, and let $R_i \in \{1,\dots,K\}$ denote the unobserved race of individual $i$ in the sample. Let $G_i$ be their Census geography, and $N_i$ be their surname. BISG estimates the posterior probability $\p(R_i = r \mid N_i, G_i)$ using Bayes’ rule. The geographic prior is given by $\p(R_i = r \mid G_i = g)$, constructed from Census counts in geography $g$. Name information is incorporated through $\p(N_i = n \mid R_i = r)$, estimated from Census surname-race tables. Under the standard assumption that surname and geography are independent given race, the posterior is
\[ \p(R_i = r \mid N_i = n, G_i = g) = \frac{\p(N_i = n \mid R_i = r)\p(R_i = r \mid G_i = g)}{\sum_{r'=1}^K \p(N_i = n \mid R_i = r')\p(R_i = r' \mid G_i = g)}. \]

BISG has been widely used in social science to study the disparate racial impacts of key policies, in fields such as policing \citep{edwa:lee:espo:19}, lending \citep{zhan:18}, and eviction \citep{hepb:loui:desm:20}. Several improvements to BISG have been suggested over the years, including relaxing key independence assumptions \citep{McCartan09092025} and calibrating predictions to known totals \cite{10.1093/jrsssa/qnaf003}. We utilize an improved version introduced in \cite{imai2022addressing}, which allows for incorporation of first and middle names into the predictions. This version also expands the dictionaries of race-name distributions used to make predictions \citep{imai2022addressing}. The enhanced version is implemented in the \texttt{R} software package \texttt{WRU} \citep{wru}.

\section{Ecological Inference}\label{sec:ecoreg}

We utilize the modern ecological inference method of \cite{fishman2024estimating}. This approach models individual-level vote choices as independent Bernoulli variables, with success probabilities a function of individual-level covariates. It then treats the observed aggregate vote tallies as the sum of these Bernoulli variables. 

In precinct $i$, we observe covariates $\boldsymbol{x_{ij}}$ for each individual $j\in\{1,\dots,n_i\}$ (e.g., voter-file covariates). But we do not observe the latent binary outcome $Y_{ij}$ (e.g., vote for a given candidate). Rather, we observe the aggregate count
\[ D_i \;=\; \sum_{j=1}^{n_i} Y_{ij}. \]
Using the logistic regression model, we have
\[ Y_{ij}\mid  \boldsymbol{x_{ij}} \sim \text{Bernoulli}(p_{ij}),\qquad p_{ij}=\sigma(\bbeta^\top \boldsymbol{x_{ij}}), \]
where $\bbeta$ is a coefficient vector. 

Under our independence assumption, $D_i$ follows a Poisson-Binomial distribution with parameter vector $(p_{i1},\dots,p_{in_i})$. The corresponding likelihood is 
\[ \mathcal{L}(\beta) = \prod_i \p\!\Big(D_i \,\Big|\, \{p_{ij}(\bbeta)\}_{j=1}^{n_i}\Big) = \prod_i \sum_{A\subseteq[n_i]:\,|A|=D_i}\;\prod_{j\in A}p_{ij}(\bbeta)\prod_{j\notin A}\{1-p_{ij}(\bbeta)\}, \]
where $[n_i] = \{1, 2, \dots, n_i\}$. This likelihood connects individual covariates to the observed, precinct-level vote tallies; following standard estimation procedures, our goal is to maximize it in $\bbeta$. 

For problems of our scale, the central practical obstacle is computational. The Poisson–Binomial probability mass function is computationally infeasible to evaluate because it sums over $\binom{n_i}{D_i}$ subsets of the voters, which grows extremely rapidly with precinct size. Our method uses the Lyapunov CLT to approximate the $D_i$ as Gaussians whose mean and variance match those of the corresponding Poisson-Binomials random variables. This yields computable gradients and fast training for Poisson-Binomial GLMs at large scale. This method was discussed in \cite{RosenmanViswanathan2018} and \cite{Rosenman2019}, and formalized in \cite{fishman2024estimating}. Similar methods have also been deployed in the econometric literature \citep{Ainsworth2020, BerryCoxHaile2025}.

In our case, we use a standardized set of covariates for fitting each model. These include each voter's: 
\begin{itemize}
\item party (Democrat, Republican, or Independent),
\item gender (male or female),
\item age (discretized by decade),
\item race predictions (predicted probabilities of being White, Black, Hispanic, or Asian, from \texttt{wru}), 
\item multifamily housing status (0 or 1; inferred from the individual's address),
\item ballot type in the predicted election (voted in person or voted by mail),
\item participation in the prior presidential election (0 or 1), and 
\item participation in the prior midterm election (0 or 1).
\end{itemize}

While the approximate log-likelihood is not guaranteed to be concave \citep{Rosenman2019}, we find good convergence properties from all ecological models using this set of covariates. For a review of the full set of assumptions underlying modern ecological inference techniques, see \cite{kuriwaki2026role}. 

\section{District-Specific Electoral Racial Compositions}\label{sec:distRaceComp}

\begin{figure}[H]
\centering
\includegraphics[width=\textwidth]{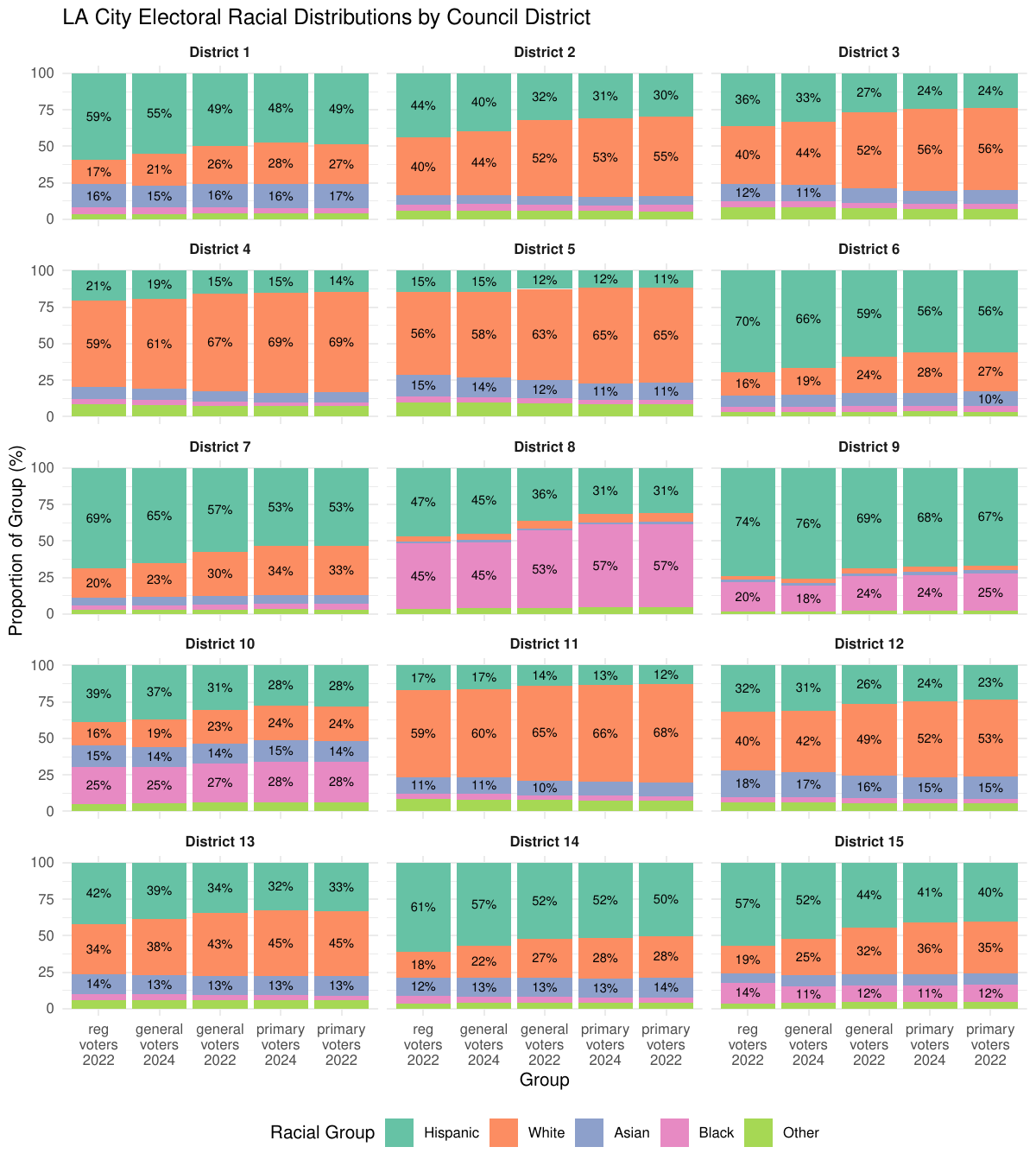}
\caption{
Estimated racial composition of registered voters and 2022 and 2024 electorates within each of L.A.'s 15 City Council districts. Race is imputed using Bayesian Improved Surname Geocoding (BISG). For readability, labels are provided only for groups that comprise 10\% or more of each group.
}
\label{fig:LARacialByDistrict}
\end{figure}

\newpage
\section{Proration from Census blocks to precincts}\label{sec:prorate}

In order to perform our analysis, we needed all of our data at the same level of granularity.  Census population data was available at the census block level while elections results were available only for each voting precinct.  While some census blocks nest entirely within a precinct, many do not, and we had to decide how to assign population values for these census blocks to the appropriate precincts.  Standard methods~\cite{maup} simply assign all the population within a census block to the precinct with which it shares the most area of overlap, but we took a more careful approach. 

The L.A. voter file was obtained from L2, Inc., a leading nonpartisan data vendor. Each address in the L2 file was geocoded to a latitude and longitude, and matched to the relevant Census block.
We were then able to use this voter data to prorate population.  For example, if Census Block 1 has 10 voters, and it overlaps Precinct A and Precinct B with 6 of its voters in precinct A and 4 of its voters in Precinct B, then 60\% of Census Block 1's population was prorated to Precinct A and 40\% of its population was prorated to Precinct B.  There were a small number of voters located in a census block not in L.A. or in a census block and a voting precinct that did not overlap (likely due to geocoding errors or voters moving), and these were not considered in the proration.  Finally, there were a small number of census blocks in which there was no voter information, and these were prorated using previous approaches, assigning population to the precinct with which it shares the most area.

\section{Creating Ensembles and the Recombination Markov chain}\label{sec:recom}

Throughout, our random sampling of possible districting plans is done using the Recombination Markov chain~\cite{DeFord2021Recombination}, as implemented in the python package {\it GerryChain}~\cite{gerrychain}. This is a random process that starts at any districting plan, randomly chooses two districts, merges them together, draws a spanning tree of this union, picks a random edge of this spanning tree, 
and if this edge splits the tree into two pieces with approximately equal population such a split is performed to produce two new districts. 
This process is repeated a large number of times, to produce a large number of possible districting plans.  We use starting districting plans randomly chosen with gerrychain's \texttt{recursive\_tree\_part()} function. We perform 300,000 steps of this process for each possible number of districts we consider; see the next section for justification of this choice. 

We use a population tolerance of $\varepsilon = 0.05$, meaning we require districts to always remain within $\pm 5\%$ of their ideal population, which we define as L.A.'s total population divided by the number of districts.  This population tolerance is consistent with the enacted plan, which shows a maximum deviation from districts' ideal population of just under $5\%$. For each spanning tree, we try 10 times to pick a random edge and see if it produces an approximately balanced split before giving up and starting the process over.

\section{Convergence Checks for the Recombination Markov chain}\label{sec:mixing}

\begin{figure}[H]
\centering

\subfloat[]{%
\begin{minipage}{.49\textwidth}
\centering
\includegraphics[width=\linewidth]{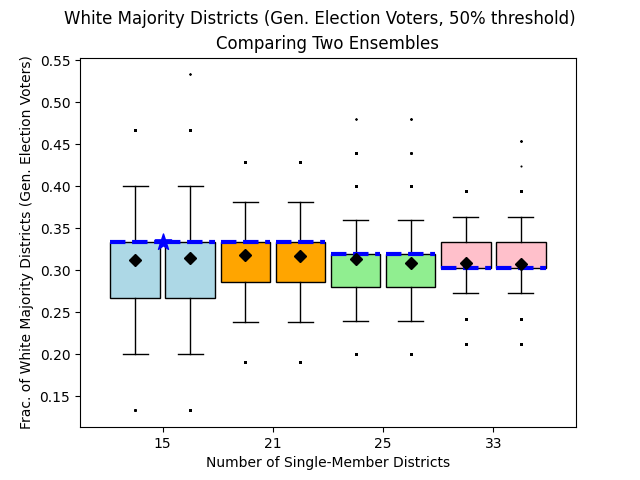}
\label{fig:WGEN_conv}
\end{minipage}
}
\hfill
\subfloat[]{%
\begin{minipage}{.49\textwidth}
\centering
\includegraphics[width=\linewidth]{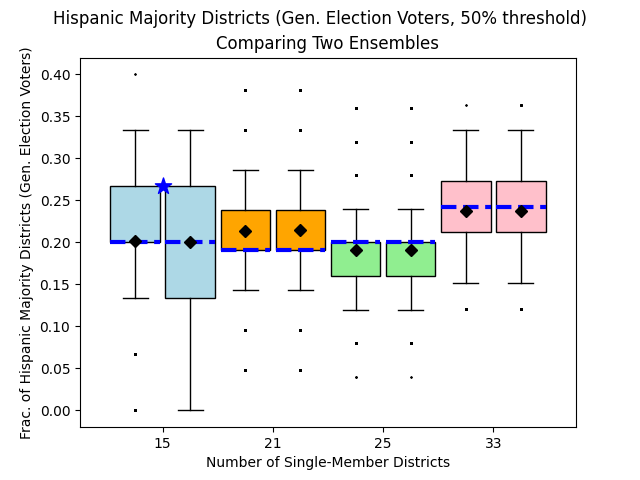}
\label{fig:HGEN_conv}
\end{minipage}
}

\vspace{0.5em}

\subfloat[]{%
\begin{minipage}{.49\textwidth}
\centering
\includegraphics[width=\linewidth]{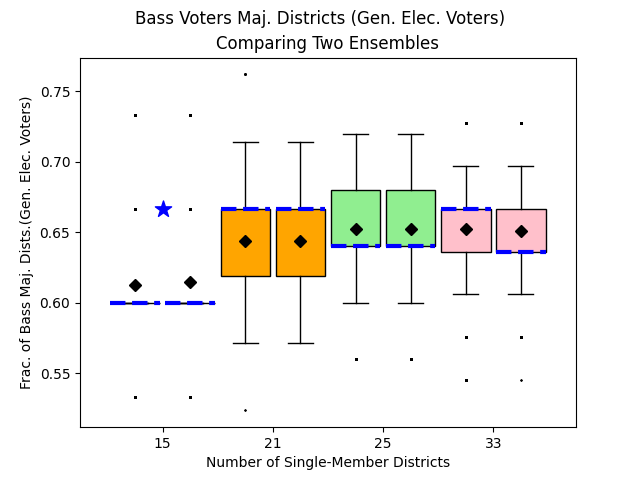}
\label{fig:Bass_conv}
\end{minipage}
}
\hfill
\subfloat[]{%
\begin{minipage}{.49\textwidth}
\centering
\includegraphics[width=\linewidth]{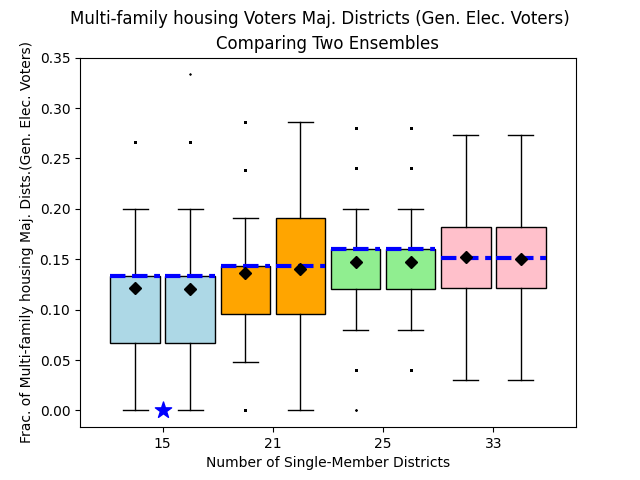}
\label{fig:MGEN_conv}
\end{minipage}
}

\caption{
For all quantities considered in Figure~3, we show the corresponding distributions obtained from an independent Markov chain run of length 300,000 steps. The resulting distributions are nearly identical to those shown in Figure~3, providing evidence that a chain length of 300,000 steps is sufficient for the quantities considered here.
}
\label{fig:W&HGEN_conv}
\end{figure}

\begin{figure}[H]
\centering
\subfloat[]{%
\begin{minipage}{.49\textwidth}
    \centering
    \includegraphics[width=1\linewidth]{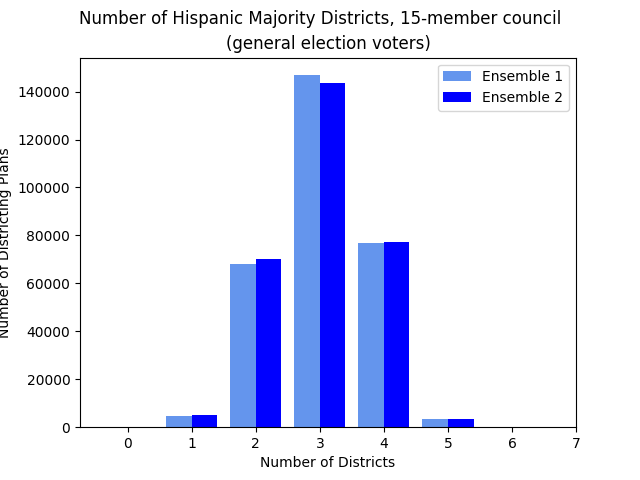}
    \label{fig:conv_HGEN15}

\end{minipage}%
}\subfloat[]{%
\begin{minipage}{.49\textwidth}
    \centering
    \includegraphics[width=1\linewidth]{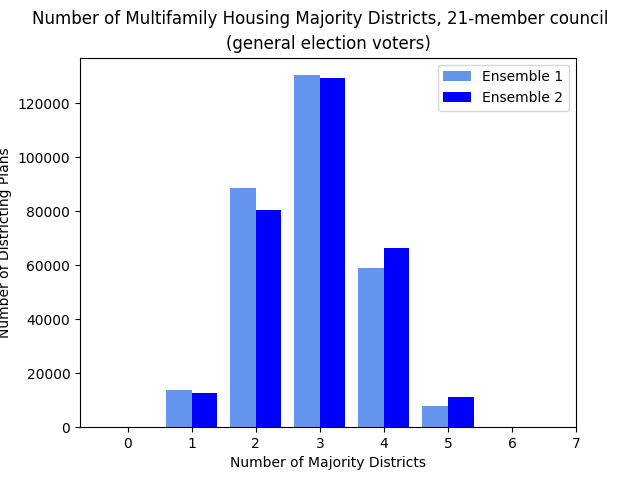}
    \label{fig:conv_MGEN21}

\end{minipage}%
}
\caption{For two boxplots in the previous figure whose quartiles don't match, the distributions of majority districts are extremely close. (a) Hispanic majority districts when there's 15 districts; (b) Multifamily Housing majority districts when there's 21 districts. }
\label{fig:convergence_hists}
\end{figure}

If ReCom is run for too few steps, we risk our randomly generated plans being too similar to the starting state, thus biasing our conclusions.  To be sure the process has run for long enough, we perform two different runs from different starting states and compare the results.  If resulting plots look noticeably different, this is a sign that our Markov chain has not run for enough steps.  After seeing discrepancies after 100,000 and 200,000 steps, we settled on 300,000 steps as likely sufficient for our purposes.  While we examined many statistics to check convergence, we include in Figure~\ref{fig:W&HGEN_conv} the plots for all statistics evaluated in Figure~3. In Figure~\ref{fig:convergence_hists}, we include histograms of the quantities for which the quartiles in Figure~\ref{fig:W&HGEN_conv} don't perfectly match, namely Hispanic-majority districts when there are 15 districts and multifamily housing majority districts when there are 21 districts.  We see that despite the quartiles differing by one district, the histograms show the distributions of these quantities across both ensembles are extremely similar. Histograms of other quantities looked similar to those in Figure~\ref{fig:convergence_hists}. 

\begin{figure}[H]
    \centering
    \includegraphics[width=0.7\linewidth]{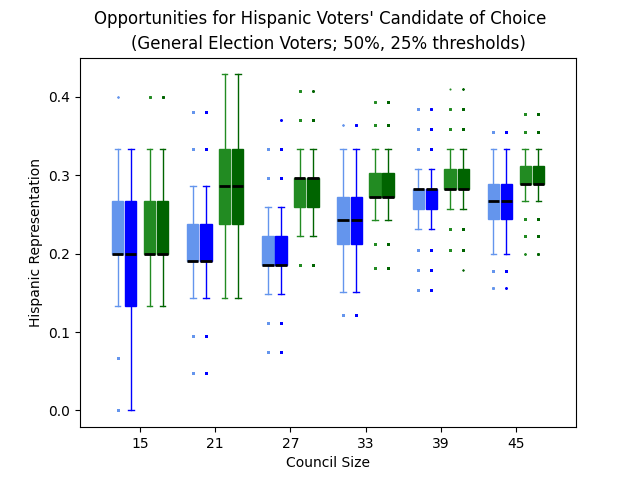}
    \caption{
    We compare the results of two ensembles, one given by light blue (single-member districts) and light green (multimember districts), and the other given by dark blue (single-member districts) and dark green (multimember districts). We see the two ensembles give nearly identical results for Hispanic Representation, suggesting the 300,000 steps performed for each suffices. (see also Figure~5) }
    \label{fig:HGEN_50p_25p_conv}
\end{figure}

In Section~4, we considered a broader set of ensembles, looking at both single-member districts and three-member districts for Councils with 15, 21, 27, 33, 39, and 45 members. We examined many statistics to check for convergence, and saw similar results suggesting 300,000 steps was sufficient in both the cases.  For example, in Figure~\ref{fig:HGEN_50p_25p_conv}, we present the results of two different ensembles and see that no matter which ensemble is used, the results are nearly identical to those originally presented Figure~5. Histograms of individual statistics, for both single-member and multimember districts, showed discrepancies between the two ensembles comparable to those shown in Figure~\ref{fig:convergence_hists}.

\section{Additional Plots for Single-member Districts}
\label{sec:more_single_plots}

Here we include additional plots looking at changes in representation for different numbers of single-member districts, in addition to those for White-majority, Hispanic-majority, Bass voter-majority, and multifamily housing majority districts shown in Figure~3. 

\begin{figure}[H]
\centering
\subfloat[]{%
\begin{minipage}{.49\textwidth}
    \centering
    \includegraphics[width=1\linewidth]{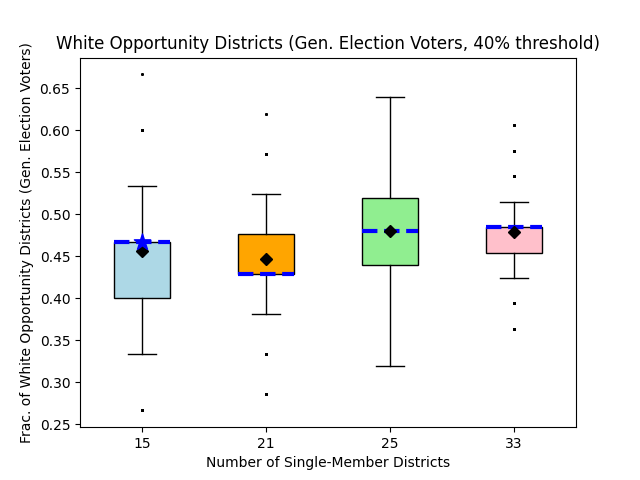}
    \label{fig:WGEN_40p}

\end{minipage}%
}\subfloat[]{%
\begin{minipage}{.49\textwidth}
    \centering
    \includegraphics[width=1\linewidth]{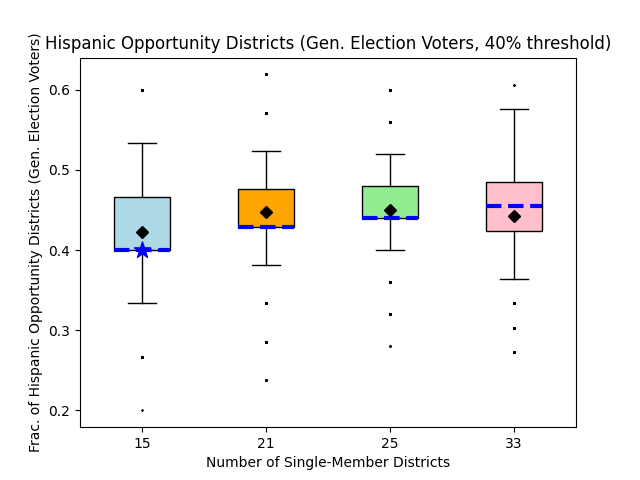}
    \label{fig:HGEN_40p}

\end{minipage}%
}
    \caption{When considering opportunity districts (where a group comprises at least 40\% or a district) rather than majority districts (where a group comprises at least 50\% of a district), there are still few changes when the number of districts changes. }
    \label{fig:opp}
\end{figure}

Often a group does not need to comprise 50\% of a district to have an opportunity to elect a representative of their choice. Because of this this, we also considered White opportunity districts and Hispanic opportunity districts, which we define as comprising more than 40\% of a district. The fraction of White and Hispanic opportunity districts are shown in Figure~\ref{fig:opp}. Again, there is little change when the number of districts change, illustrating that increasing the number of Council districts will have little impact on opportunities for racial representation. 

\begin{figure}[H]
\centering
\subfloat[]{%
\begin{minipage}{.49\textwidth}
    \centering
    \includegraphics[width=1\linewidth]{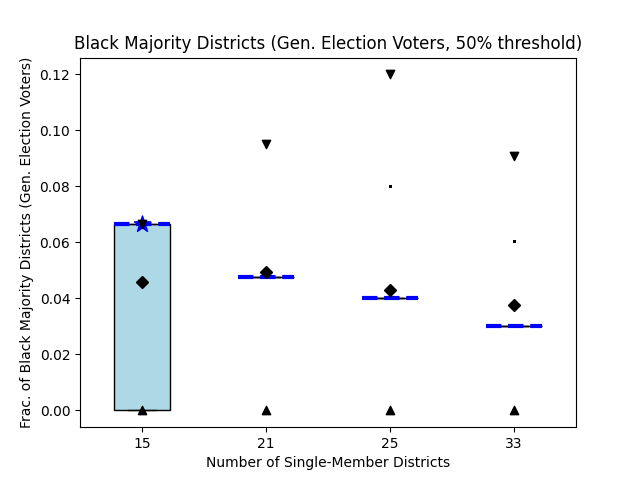}
    \label{fig:BGEN}

\end{minipage}%
}\subfloat[]{%
\begin{minipage}{.49\textwidth}
    \centering
    \includegraphics[width=1\linewidth]{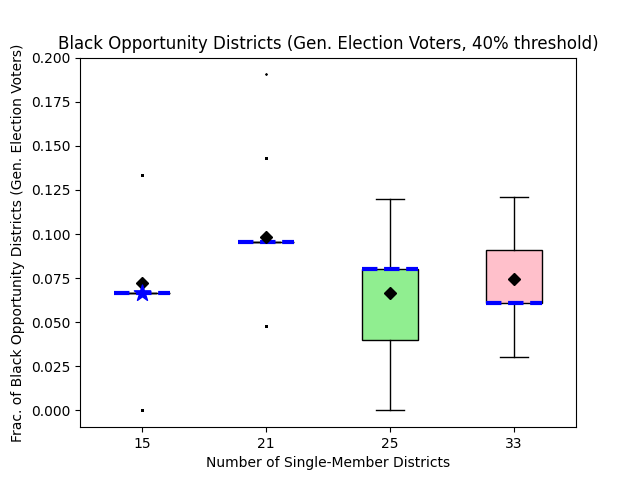}
    \label{fig:BGEN_40p}

\end{minipage}%
}
    \caption{(a) The fraction of Black majority districts across different Council sizes. (b) The fraction of Black opportunity districts ($\geq 40\%$ Black) across different Council sizes.}
    \label{fig:black}
\end{figure}

Beyond White and Hispanic representation, we note that no plans in our ensemble had any Asian majority or Other majority districts, when there were 15, 21, 25, or 33 districts; the same is true for Asian-opportunity and Other-opportunity districts calculated using a 40\% threshold. 
Figure~\ref{fig:black} considers Black-majority and Black-opportunity districts. We note that while there appears to be a bit more variation across different Council sizes, this is largely an artifact of the small numbers and that for $d$ districts, only the values $0/d$, $1/d$, $2/d$, etc. are possible values for the fraction of districts that are Black majority/opportunity districts.

\begin{figure}[H]
\centering
\subfloat[]{%
\begin{minipage}{.49\textwidth}
    \centering
    \includegraphics[width=1\linewidth]{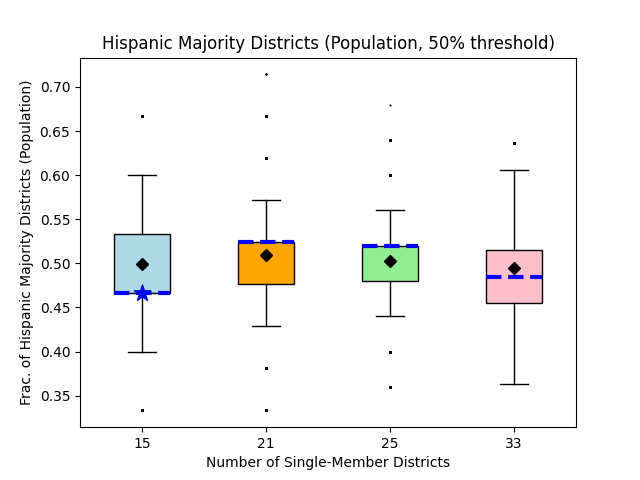}
    \label{fig:HPOP}

\end{minipage}%
}\subfloat[]{%
\begin{minipage}{.49\textwidth}
    \centering
    \includegraphics[width=1\linewidth]{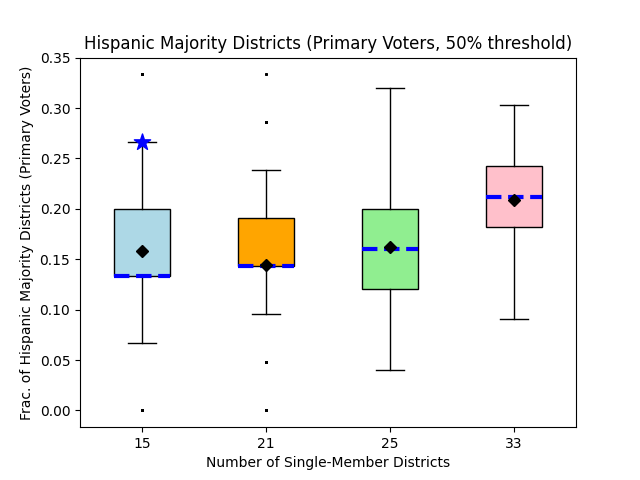}
    \label{fig:HPRI}

\end{minipage}%
}
    \caption{(a) The fraction of districts for which the population is majority-Hispanic    
    across different Council sizes. (b) The fraction districts for which the primary voters are majority-Hispanic across different Council sizes.}
    \label{fig:HPOP-HPRI}
\end{figure}

\begin{figure}[H]
\centering
\subfloat[]{%
\begin{minipage}{.49\textwidth}
    \centering
    \includegraphics[width=1\linewidth]{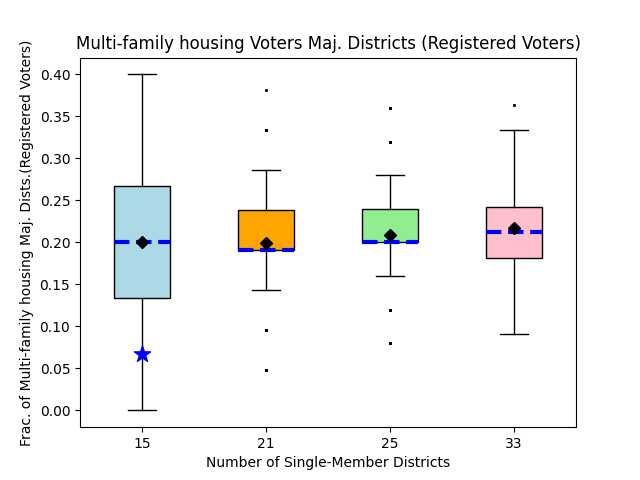}
    \label{fig:MREG}

\end{minipage}%
}\subfloat[]{%
\begin{minipage}{.49\textwidth}
    \centering
    \includegraphics[width=1\linewidth]{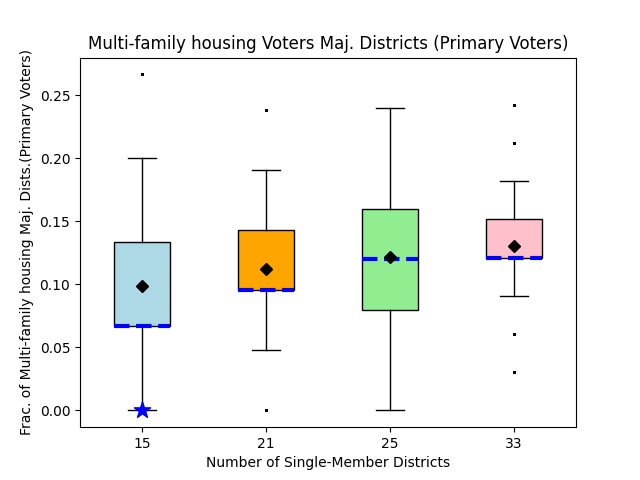}
    \label{fig:MPRI}

\end{minipage}%
}
    \caption{(a) The fraction of districts for which a majority of the registered voters live in multifamily housing
    across different Council sizes. (b) The fraction districts for a majority of the primary voters like in multifamily housing across different Council sizes.}
    \label{fig:MREG-MPRI}
\end{figure}

All of our analysis above has focused on general election voters, as that is the most relevant group to consider for studying potential outcomes of future elections.  However, we see the same pattern when looking at other groups, such as population or primary voters. In Figure~\ref{fig:HPOP-HPRI}, we see there is little change across council sizes for the fraction of districts that are Hispanic-majority with respect to population and the fraction of districts that are Hispanic-majority with respect to primary voters (though these quantities are themselves different from each other). 
Similarly, Figure~\ref{fig:MREG-MPRI} considers the fraction of districts where a majority of the registered voters live in multifamily housing and the fraction of districts where a majority of the primary voters live in multifamily housing, and we see each remains largely unchanged when the overall number of districts changes.

Many other figures examining these quantities or combinations thereof are available upon request. All show a similar pattern of little change across different Council sizes.

\section{Short Bursts}\label{sec:sb}

In addition to exploring a wide range of districting plans, from the typical to the merely possible, computational methods can allow us to find potential plans that optimize certain qualities (such as minority representation) while still meeting baseline criteria regarding compactness and population equality. 
To that point, we employ the ``short bursts" technique
to explore extreme cases~\cite{shortbursts}. For instance, we can use this method to look for the greatest and least possible numbers of White and Hispanic-majority districts. 

In the short bursts method, a Markov chain such a ReCom is run for some small number of steps, in our case 10 steps, which we call a {\it short burst}. We choose the plan from among those 10 that has the best value of a statistic we're interested in (for example, the plan with the most Hispanic-majority districts). We then restart from that plan for another 10 steps and repeat. Overall, we performed 30,000 bursts of length 10 for a total of 300,000 steps, and did this independently for each number of districts, for each quantity we wanted to optimize, and for each quantity we wanted to minimize. We also repeated each experiment twice. We tracked the maximum or minimum found, and these were plotted as triangles in the appropriate figures (Figures~3a, 3b, 5, 6, \ref{fig:BGEN}, \ref{fig:TBass_50p_25p}, and \ref{fig:TPadilla_50p_25p}). We note that while this method attempts to maximize or minimize certain quantities and has been shown to outperform other methods in doing do, we are not able to guaranteed that we have found the maximum or minimum possible values.

\section{Creating Ranked Ballots from Unranked Voting Data}\label{sec:ranked_ballots}

To create sample ranked ballots, we used the Cambridge sampler developed through the \href{https://github.com/mggg/VoteKit}{Votekit API} to generate ballots of varying lengths. 
The Cambridge Sampler creates ballots ranking candidates from two blocs (in our case, progressive and moderate) based on ballot behavior (including shortened ballots) from two-slate ranked choice elections in Cambridge, MA, which is one of the only municipalities in the U.S. to continuously employ ranked-choice voting in their City Council over many years. This sampler uses the Placket-Luce model, and the slate of six candidates were categorized into either a progressive or moderate bloc. The three candidates in the progressive bloc were Bass, Feuer, and Viola. the three candidates in the moderate bloc were Caruso, De León, and Kim. The parameters were created by using data from the 2022 Primary Election for each precinct. The cohesion parameters were set to 0.75 for the corresponding bloc to closest resemble relative cohesion within a slate. These values represent the degree of unity that voters for each bloc vote in. The bloc ratios were calculated from the number of votes per each candidate for each precinct on the primary data. Alpha, which demonstrates the spread/distribution of voters, was set to 0.9 for cohesion within each bloc, allowing some crossover voting while ensuring general relative cohesion. Overall bloc proportion was calculate by normalizing the ratio of general 2022 election votes for Caruso and Bass. 

Most ballots had less than six candidates ranked, and many stuck to their own slate. There were a few that ranked all candidates, but those were negligible compared to the number ranking merely the candidates in their own bloc, which was dependent on the value of alpha.

\section{Simulating IRV and STV elections}\label{sec:irv-stv}

After the creation of the ballots, we run Instant Runoff Voting on each potential plan, aggregating the winners of each district over all district plans of 15, 21, 25, and 33 districts. By performing 300,000 steps of ReCom, we ensure that there is enough variation that the resulting map the results can be independent of any anomalies in any single plan and all starting states randomly chosen. Each run of the chain use the same simulated ballots in order to focus on the redistricting effect. Both of the runs produced similar results, ensuring that 300,000 steps is enough to eliminate any potential dependencies.
We aggregate the data over 300,000 different assignments in order to ensure accuracy, and ran walks twice to verify our results.

For multimember districts, we run Single Transferrable Voting on each plan, aggregating three winners for each district of the 5 districts plan. We perform 10 walks of 30,000 steps and aggregate the results, using the same Markov Chain Monte Carlo method.

\begin{figure}[H]
    \centering
    \includegraphics[width=1\linewidth]{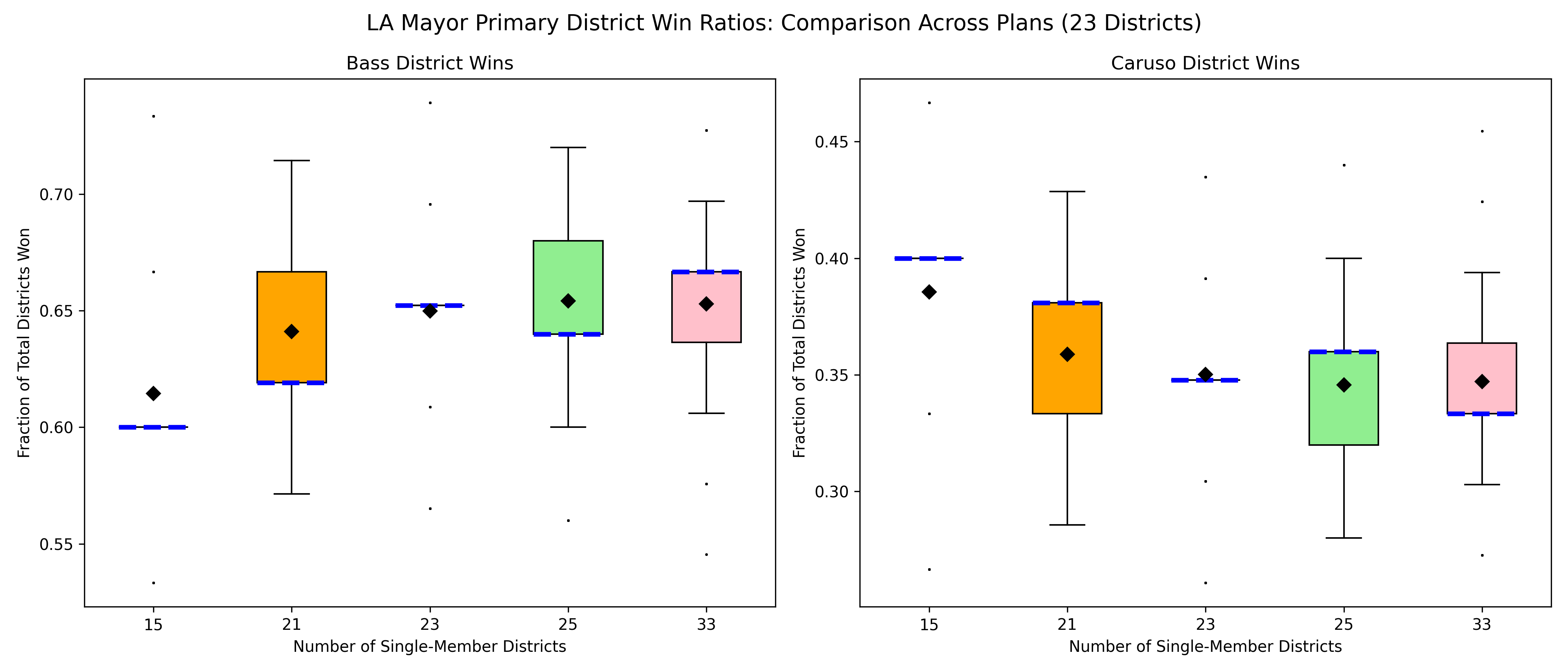}
    \caption{Distribution of districts won by Bass and Caruso per 15, 21, 23, 25, and 33 district plan. The colored boxes are the middle 50\% of the districts won per candidate, the black diamonds are the ensemble means, and the blue lines are the median values of districts won.}
    \label{fig:irv-results-23}
\end{figure}

\section{Additional Plots Comparing Single-member and Three-member Districts}
\label{sec:single_vs_multi_figs}

\begin{figure}[H]
    \centering
    \includegraphics[width=0.7\linewidth]{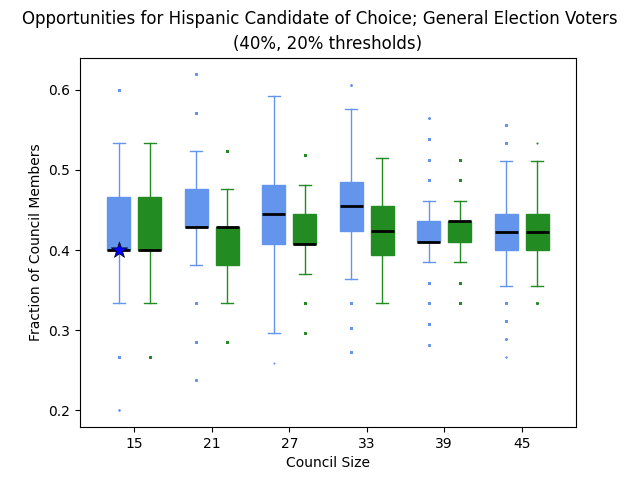}
    \caption{
    Proportion of candidates that could be picked by Hispanic voters using 40\% thresholds in single-member districts (in blue) and 20\% thresholds in three-member districts (in green). Similar to Figure~5 which uses 50\% and 25\% thresholds, three-member districts provides more opportunities for the candidates preferred by Hispanic voters, with little change across the Council size.  }
    \label{fig:HGEN_40p_20p}
\end{figure}

Often a group does not need to meet the strict thresholds we define of 50\% in single-member districts and 25\% in multimember districts to elect their representative of choice, as candidates often receive additional votes from other groups. Because of this this, we also considered Hispanic opportunity districts, which we define as comprising more than 40\% of a single-member district or more than 20\% of a multimember district (with the opportunity to elect two candidates of choice if they comprise more than 40\% of a multimember district, and the opportunity to elect three candidates of choice if they comprise more than 60\% of a multimember district). The fraction of representatives the Hispanic community would have the opportunity to choose under these assumptions are shown in Figure~\ref{fig:HGEN_40p_20p}, where the picture is similar to Figure~5 which uses the stricter 50\% and 25\% thresholds, respectively.

\begin{figure}[H]
    \centering
    \includegraphics[width=0.7\linewidth]{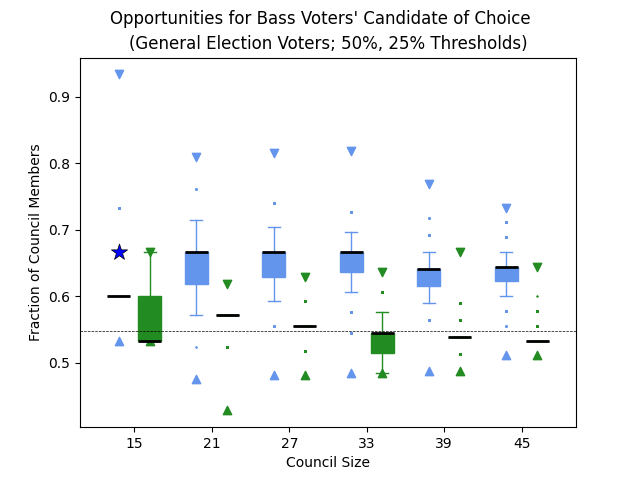}
    \caption{Proportion of candidates that could be picked by voters preferring Bass in the 2022 Mayoral election, for single-member districts (in blue) and three-member districts (in green). Median values are given by thick horizontal black lines. The horizontal dashed line stretching the length of the plot gives the overall fraction of votes for Bass, a target if proportionality of election outcomes is a goal. }
    \label{fig:TBass_50p_25p}
\end{figure}

\begin{figure}[H]
    \centering
    \includegraphics[width=0.7\linewidth]{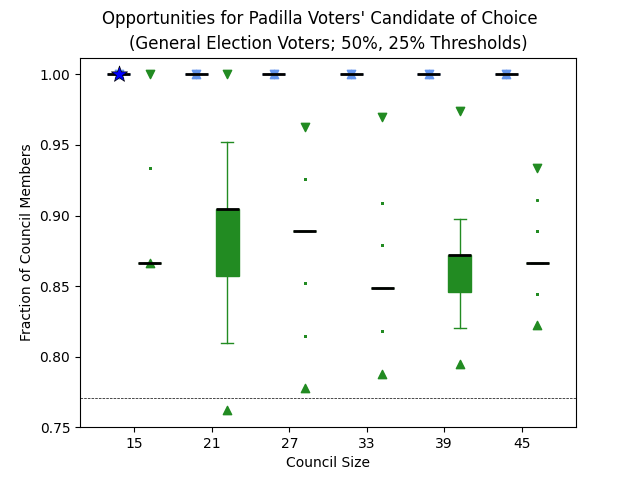}
    \caption{Proportion of candidates that could be picked by voters preferring Padilla in the 2022 U.S. Senate election, for single-member districts (in blue) and three-member districts (in green). Median values are given by thick horizontal black lines. The horizontal dashed line stretching the length of the plot gives the overall fraction of votes for Padilla, a target if proportionality of election outcomes is a goal.}
    \label{fig:TPadilla_50p_25p}
\end{figure}

Beyond racial representation (and representation based on housing status, as we considered in Figure~6), we also considered political representation. Figures~\ref{fig:TBass_50p_25p} and~\ref{fig:TPadilla_50p_25p} show that when considering votes cast in the 2022 Mayoral and U.S. Senate races, respectively, three-member districts make it possible for outcomes to more closely reflect the opinions of voters.

\end{document}